\documentclass[12pt,preprint]{aastex}

\shorttitle{X-ray Optical Depth in Stellar Coronae}
\shortauthors{Testa et al.}

\def \lya  {Ly$\alpha$}
\def \lyb  {Ly$\beta$}
\def \lyab {Ly$\alpha$/Ly$\beta$}
\def \lyba {Ly$\beta$/Ly$\alpha$}

\def \lx   {$L_{\rm X}$}

\def \Lx   {$L_{\rm X}$}
\def \lbol {$L_{\rm bol}$}

\def \ta  {Paper~I}
\def \tb  {TDP04}

\def \cha   {{\em Chandra}}
\def \hetgs {{\sc hetgs}}
\def \hetg  {{\sc hetg}}
\def \heg  {{\sc heg}}
\def \meg  {{\sc meg}}

\def \fexvii  {Fe\,{\sc xvii}}
\def \fexviii {Fe\,{\sc xviii}}
\def \fexxi  {Fe\,{\sc xxi}}
\def \neix   {Ne\,{\sc ix}}
\def \nex    {Ne\,{\sc x}}
\def \ovii   {O\,{\sc vii}}
\def \oviii  {O\,{\sc viii}}
\def \mgxi   {Mg\,{\sc xi}}

\begin{document}
\title{On X-ray Optical Depth in the Coronae of Active Stars}
\author{Paola Testa\altaffilmark{1}, Jeremy
J. Drake\altaffilmark{2}, Giovanni Peres\altaffilmark{3},
David P. Huenemoerder\altaffilmark{1}}
\altaffiltext{1}{Massachusetts Institute of Technology, Kavli 
Institute for Astrophysics and Space Research, 70 Vassar street, 
Cambridge, MA 02139, USA; testa@space.mit.edu}
\altaffiltext{2}{Smithsonian Astrophysical Observatory, MS 
3, 60 Garden Street, Cambridge, MA 02138, USA}
\altaffiltext{3}{Dipartimento di Scienze Fisiche \& Astronomiche, 
Sezione di Astronomia, Universit\`a di Palermo Piazza del Parlamento 
1, 90134 Palermo, Italy}

\begin{abstract}
We have investigated the optical thickness of the coronal plasma 
through the analysis of high-resolution X-ray spectra of a large 
sample of active stars observed with the High Energy Transmission 
Grating Spectrometer on {\em Chandra}. In particular, we probed for 
the presence of significant resonant scattering in the strong Lyman 
series lines arising from hydrogen-like oxygen and neon ions.
The active RS CVn-type binaries II~Peg and IM~Peg and the single M 
dwarf EV~Lac show significant optical depth. For these active coronae, 
the \lyab\ ratios are significantly depleted as compared with 
theoretical predictions and with the same ratios observed in similar 
active stars.  Interpreting these decrements in terms of resonance 
scattering of line photons out of the line-of-sight, we are able to 
derive an estimate for the typical size of coronal structures, and 
from these we also derive estimates of coronal filling factors. For 
all three sources we find that the both the photon path length as a 
fraction of the stellar radius, and the implied surface filling 
factors are very small and amount to a few percent at most.  
The measured \lyab\ ratios are in good agreement with APED theoretical
predictions, thus indicating negligible optical depth, for the other 
sources in our sample. 
We discuss the implications for coronal structuring and heating flux 
requirements.   
For the stellar sample as a whole, the data suggest increasing 
quenching of \lya\ relative to \lyb\ as function of both \lx/\lbol\ 
and the density-sensitive \mgxi\ forbidden to intercombination line 
ratio, as might generally be expected. 
\end{abstract}


\section{Introduction}
\label{s:intro}

A fundamental issue in the physics of stellar outer atmospheres 
concerns the relationship between magnetic activity on stars with
a wide range of physical parameters and solar magnetic activity.  
How directly and how far does the solar analogy apply to other stars, 
and do any of the underlying physical processes differ?  The X-ray
luminosities of late-type stars can span several decades 
\citep[e.g.,][]{Vaiana81}, and these hot coronae are found on such a 
wide range of spectral types that the extrapolation of the now 
well-studied solar corona to the extremes of stellar activity is by no
means obvious and could be inappropriate.  
Indeed, ``scaled up Sun'' scenarios, in which a stellar surface is 
covered with bright solar-like active regions, only realise X-ray 
luminosities 100 times that of the typical active Sun 
\citep[e.g.,][]{Drake00}.  The most active stars, with X-ray 
luminosities of up to 10,000 times the solar X-ray luminosity, must 
have coronae which are structured differently in some way.

Since coronal structures can be imaged presently only on the Sun, the 
structuring of other stellar coronae is generally investigated through 
the application of techniques such as the study of lightcurves during 
flares \citep[e.g.,][]{Schmitt99,Favata00,Maggio00,Reale04,Testa07}, 
rotational modulation \citep[e.g.,][]{Brickhouse01,MarinoL03,Huenemoerder06}, 
and study of density properties together with information on emission 
measure \citep[e.g.,][ hereafter \tb; \citealt{Ness04}]{Testa04b}.
Most of these analyses indicate that the emitting plasma is rather 
compact (scale height $\leq 0.5 R_{\star}$) and localized at high 
latitude \citep[see e.g.,][ \tb]{Schmitt99,Brickhouse01,Testa04a}; 
however, the presence of extended coronal plasma has also 
been claimed on some stars based on UV and X-ray Doppler studies 
\citep[e.g.][]{Chung04,Redfield03}.

The search for signs of quenching in strong lines through resonance 
scattering represents a further technique that offers a potentially 
powerful diagnostic of the sizes of X-ray emitting regions; the 
escape probability of a photon emitted by a resonance line in a 
low density homogeneous plasma is in fact dependent on the 
line-of-sight path length through the plasma region.  
Significant scattering optical depth can be combined with density 
measurements to obtain an estimate of photon path length within the
emitting plasma.

Several existing studies have explored optical depths of both solar 
and stellar coronal emission lines.  Studies of solar X-ray spectra
have aimed at probing the optical depth in the strong ($gf=2.66$) 
$2p^5 3d ^1P_1 - 2p^6\, ^1S_0$ resonance line of \fexvii\ at 
15.01~\AA\ as compared to nearby weaker \fexvii\ lines, though with 
controversial results concerning whether optical depth effects were 
seen or not \citep{Phillips96,Phillips97,Schmelz97,Saba99}.
In particular, \citet{Saba99} review recent observational findings
on the opacity inferred from the study of the bright iron resonance 
line at 15.01~\AA\ and on the center-to-limb behaviour. 
Among other issues, \citet{Saba99} address the discrepancy they find 
in the derived direction and magnitude of the center-to-limb trend 
(also in agreement with \citealt{Schmelz97}), as compared to the 
findings of \citet{Phillips96} who find that the effect of resonant 
scattering is decreasing from the disk center toward the solar limb, 
a trend irreconcilable and totally opposite to that found by 
\citet{Saba99} and \citet{Schmelz97}. \citet{Brickhouse06} have 
recently reanalized solar X-ray spectra and suggest that previously 
ignored blends might explain the departure of measured ratios from 
theoretical calculations.

Resonant scattering in stellar coronae has been investigated by 
\citet{Phillips01} and \citet{Ness03b} through the analysis of the 
same transition observed at high resolution ($\lambda /\Delta\lambda$ 
up to $\sim 1000$) by {\em Chandra} and XMM-{\em Newton}. 
Both stellar studies of \fexvii\ transitions fail to find evidence for 
significant deviation from the optically thin regime, and in 
particular the large survey of stellar spectra analyzed by 
\citet{Ness03b} show that no firm results can be obtained from Fe 
lines.  One exception is the suggestion of resonance scattering in
\fexvii\ 15.01~\AA\ on the basis of the observed variability of line 
ratios seen during a flare on AB Dor by \citet{Matranga05}.
We note that \citet{Phillips01} attempted to derive constraints on 
emitting region size based on their upper limit to optical depth in 
the coronae of Capella, though, as we discuss in this paper 
(\S\ref{s:discuss}), such upper limits cannot be reliably used in this
fashion because scattering {\em into} the line of sight renders 
optical depth measurements themselves only lower limits to the true 
scattering depth.

In a previous {\em Letter} \citep[][ hereafter \ta]{Testa04a} we 
presented results obtained using a different approach to the study of 
coronal optical depth, through the analysis of Ne and O \lya\ to 
\lyb\ line strength ratios as observed by the High Energy Transmission 
Grating (HETG) on board the {\it Chandra} X-ray Observatory.
Significant depletion of \lya\ lines to resonance scattering were seen
in the spectra of the RS~CVn-type binaries II~Peg and IM~Peg.
In this paper we follow up on that exploratory study and examine the 
Ne and O Lyman series lines in a large number of active stars (namely, 
the same sample for which the plasma density was analysed in \tb) in 
order to survey the X-ray optical depth properties of active stellar 
coronae.

We discuss in \S\ref{s:theor} the advantages of the Lyman series 
analysis with respect to the ``standard'' approach using \fexvii\ 
lines. 
The observations are briefly described in \S\ref{s:obs}.  Our
techniques of line flux measurement and spectral analysis are 
described in \S\ref{s:analysis}. The results are presented in 
\S\ref{ss:results}.   
We combine the results of this study with our earlier density 
estimates and discuss these in the context of coronal structure on 
active stars in \S\ref{s:discuss}; we draw our conclusions in 
\S\ref{s:conclude}.

\section{Resonant scattering in Ne and O Lyman series lines.}
	\label{s:theor}
The \fexvii\ soft X-ray complex at $\sim 15$\AA\ has been a primary
tool to probe coronal optical depth because of the large oscillator
strength of the $2p^6\, ^1S_0$-$2p^5 3d ^1P_1$ 15.01~\AA\ resonance
line and its consequent prominence in solar spectra.
However, \citet{Doron02} and \citet{Gu03} have recently shown that 
the indirect processes of radiative recombination, dielectronic 
recombination, and resonance excitation involving the neighbouring 
charge states are important for understanding the relative strengths 
of Fe\,{\sc xvii--xx} lines.  There is thus still some considerable
difficulty in reconciling theoretical and observed line strength 
ratios.  
Recently, \citet{Brickhouse06} have found good agreement of solar 
observed ratios with new theoretical calculations \citep{Chen05},
and suggest that center-to-limb observed trends 
\citep{Phillips96,Schmelz97,Saba99} are due to chance rather than
to optical depth effects. 

\begin{figure}[!ht]
\centerline{\includegraphics[width=8.5cm]{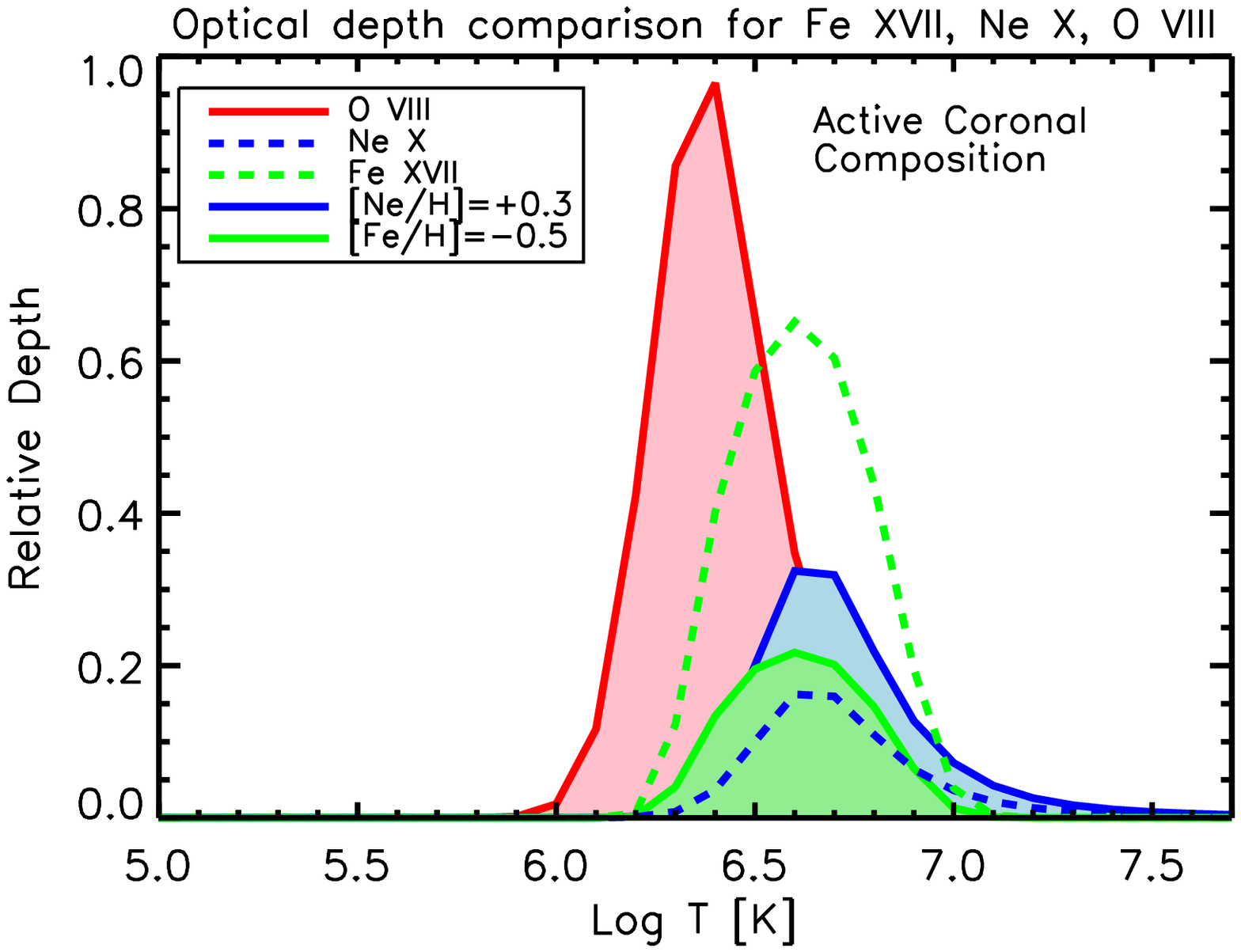}}\vspace{0.2cm}
\caption{The relative sensitivity of \lya\ lines of \nex, \oviii, and
	the strong \fexvii\ 15.01~\AA\ transition to resonant 
	scattering, as a function of plasma temperature. Relative line
	optical depths, normalised to the maximum depth in \oviii, are 
	shown for two sets of abundances. Dashed lines correspond to 
	the solar photospheric abundance mixture of 
	\citet{GrevesseSauval}; filled profiles correspond to coronal 
	abundances typically found in RS~CVn's (see text). Because of 
	the typically higher Ne abundance and lower Fe abundance found 
	in the coronae of RS~CVn systems, the expected resonant 
	scattering effects are greater for \oviii\ and \nex\ \lya\ than 
	for the \fexvii .
 	\label{fig:NeOvsFe}}
\end{figure}

An additional problem in using \fexvii\ lines as diagnostics of
optical depth is that this element has been found to be {\em depleted}
in the coronae of active stars by factors of up to 10
\citep[e.g.,][]{Drake01,Huenemoerder01,Drake03a,Audard03} as compared
with a solar or local cosmic composition. The ratio of the line-center
optical depths, $\tau_i/\tau_j$, of two lines $i$ and $j$ is given by
\begin{equation}
\frac{\tau_i}{\tau_j}=\frac{f_i \lambda_i \phi_i A_i \sqrt{m_i}}{f_j 
\lambda_j \phi_j A_j \sqrt{m_j}}
\label{e:reldep}
\end{equation}
where $f$ is the oscillator strength, $\lambda$ is the wavelength,
$\phi$ is the fractional population of the ion in question, $A$ is the
element abundance and $m$ the ion mass.  The line optical depth is 
directly proportional to abundance and, in the case of stellar 
coronae, where only the very strongest spectral lines might be 
expected to undergo any significant resonance scattering, any 
abundance depletions also reduce the sensitivity of lines as optical
depth indicators. For coronal abundances typically found in RS~CVn 
systems (see e.g.\ reviews by \citealt{Drake03a,Audard03}), we expect 
resonant lines from the more abundant ions like oxygen and neon to be 
more sensitive to resonant scattering processes than the resonant 
\fexvii\ line.  This is illustrated in Figure~\ref{fig:NeOvsFe}, where 
we show the relative optical depths for the \oviii\ and \nex\ \lya\ 
lines and the \fexvii\ $\sim 15.01$~\AA\ resonance line for both a 
representative solar chemical composition \citep{GrevesseSauval}
and for a chemical composition typically found for active stars.  For
the latter, we assumed Ne and Fe abundances 0.3 dex higher and 0.5 dex
lower, respectively, than the corresponding solar values; we note that
in several active coronae abundance anomalies even more pronounced 
have been found \citep[e.g.,][]{Brinkman01,Huenemoerder01}.  

\clearpage
\clearpage

As discussed in Paper~I, the effect of resonance scattering of \lya\
and \lyb\ photons can be diagnosed by comparison of the measured 
\lyab\ ratio with respect to the theoretical ratio.  In principle,
both $n=2 \rightarrow 1$ \lya\ and $n=3 \rightarrow 1$ \lyb\ lines 
can be affected by resonant scattering: when a large optical depth is
reached in \lya, an enhanced population of the $n=2$ level can lead 
to a potentially confusing enhancement in \lyb\ (and Ba$\alpha$) 
through collisional excitation of the $n=2\rightarrow 3$ transition. 
However, in the limit of fairly small scattering optical depth that 
we expect to characterize stellar coronae, \lyb\ essentially remains 
optically thin and should make a reliable comparison with which to 
diagnose optical depth in \lya.

\section{Observations}
\label{s:obs}

{\em Chandra} High Energy Transmission Grating Spectrometer (see
\citealt{Canizares00,hetg05} for a description of the instrumentation) 
observations of 22 cool stars covering a wide range activity level
were analysed.  The stellar parameters and particulars of the
observations were discussed in detail in a companion paper that
addressed the coronal densities of active stars (\tb).  To the sample
analysed in \tb\ we added a set of observations of IM~Peg obtained 
subsequent to the first three segments analysed in \ta\ and \tb; the 
parameters of these additional observations are listed in 
Table~\ref{tab_impeg}.

\clearpage
\begin{deluxetable}{ccccl}
\tablecolumns{4} 
\tabletypesize{\footnotesize}
\tablecaption{Parameters of the HETG observations of IM~Peg.
		 \label{tab_impeg}}
\tablewidth{0pt}
\tablehead{
 \colhead{Obs ID} & \colhead{Start date and time} & 
 \colhead{$t_{\rm exp}$}  & \colhead{\Lx \tablenotemark{a}} &  
 \colhead{\Lx \tablenotemark{b}} \\
 & & [ks] &  [erg~s$^{-1}$] & [erg~s$^{-1}$]
}
\startdata
 2527 & 2002-07-01, 15:39:08 & 24.6 & 2.95$\times 10^{31}$ & 3.04$\times 10^{31}$ \\ 
 2528 & 2002-07-08, 02:07:29 & 24.8 & 2.37$\times 10^{31}$ & 2.55$\times 10^{31}$ \\ 
 2529 & 2002-07-13, 06:27:34 & 24.8 & 2.03$\times 10^{31}$ & 2.19$\times 10^{31}$ \\
 2530 & 2002-07-18, 21:13:58 & 23.9 & 1.96$\times 10^{31}$ & 2.15$\times 10^{31}$ \\ 
 2531 & 2002-07-25, 11:30:11 & 23.9 & 1.99$\times 10^{31}$ & 2.20$\times 10^{31}$ \\ 
 2532 & 2002-07-31, 20:45:59 & 22.5 & 3.00$\times 10^{31}$ & 3.23$\times 10^{31}$ \\ 
 2533 & 2002-08-08, 00:50:22 & 23.7 & 2.57$\times 10^{31}$ & 2.79$\times 10^{31}$ \\ 
 2534 & 2002-08-15, 14:28:06 & 23.9 & 2.35$\times 10^{31}$ & 2.50$\times 10^{31}$  
 \enddata 
\tablenotetext{a}{relative to the HEG range: 1.5-15~\AA}
\tablenotetext{b}{relative to the MEG range: 2-24~\AA}
\end{deluxetable}

\clearpage

The HETG spectra and the corresponding X-ray lightcurves for the
observations can be found in \tb; we also discuss there the
variability observed for some of the stars.  The source sample covers
very different stellar characteristics, providing a wide view of X-ray
emitting stellar coronae; e.g., the X-ray luminosities (in the \hetgs\
bandpasses; see \tb) span more than five orders of magnitude, from the
relatively weak emission of Proxima Centauri with a few
$10^{26}$~erg~s$^{-1}$, up to the very high luminosity ($\sim 6 \times
10^{31}$~erg~s$^{-1}$) of the giant HD~223460.

\section{Analysis}
\label{s:analysis}

The data used here were obtained from the {\em Chandra} Data
Archive\footnote{http://cxc.harvard.edu/cda} and have been reprocessed
using standard CIAO v3.2.1 tools and analysis threads.  Effective areas
were calculated using standard CIAO procedures, which include an
appropriate observation-specific correction for the time-dependent
ACIS contamination layer.  Positive and negative spectral orders were 
summed, keeping {\sc heg} and {\sc meg} spectra separate.  For sources 
observed in several different segments, we combined the different 
observations and analysed the coadded spectra.  For IM~Peg, as noted 
above, a number of observations are available with which to probe 
optical depth properties at different times and orbital phases; 
IM~Peg will be discussed in more detail in \S\ref{sec:IMPeg}.  

Spectra were analyzed with the
PINTofALE\footnote{http://hea-www.harvard.edu/PINTofALE}
IDL\footnote{Interactive Data Language, Research Systems Inc.}
software \citep{PoA} using the technique of spectral fitting described
in \tb.  We measured the spectral line intensities of the \nex\ 
\lya\ ($2p~^2P_{3/2,1/2} - 1s~^2S_{1/2}$) and \lyb\ 
($3p~^2P_{3/2,1/2} - 1s~^2S_{1/2}$) transitions from both {\sc heg} 
and {\sc meg} spectra, and the \oviii\ lines from {\sc meg} spectra
alone (since the latter lie outside the {\sc heg} wavelength range).

In order to take into account the mild dependence of the \lyab\ ratio
on plasma temperature, we also measured the intensities of the \neix\
and \ovii\ resonance line, $r$ ($1s2p~^{1}P_{1} - 1s^{2}~^{1}S_0$ ),
providing us with an estimate of a representative temperature through
the \lya/$r$ ratio.  There are two potential problems with this
approach.  Firstly, where resonant scattering is relevant, both \lya\
and $r$ transitions can be depleted, and in such a case the \lya/$r$
ratio might deviate from its expected theoretical behaviour.
Secondly, the coronae in which these X-ray lines are formed are
expected to be characterised by continuous ranges of temperatures,
rather than by a single, isothermal plasma.  However, both of these
concerns prove unproblematic.

\begin{figure}[!ht]
\centerline{\includegraphics[width=10cm]{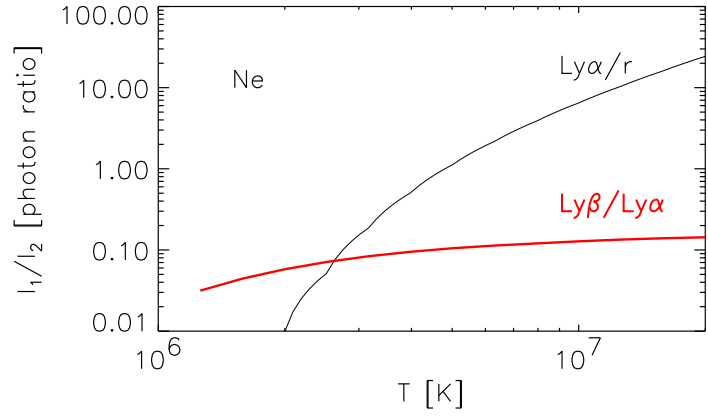}}\vspace{-0.2cm}
\caption{Temperature sensitivity (in the isothermal case) of the 
	\lya/$r$ ratio compared to the \lyba\ ratio, for neon lines. 
	The corresponding oxygen lines show analogous behaviour. 
	\label{fig:allresLyab_lumx}}
\end{figure}

Figure~\ref{fig:allresLyab_lumx} illustrates the theoretical
temperature dependence of the \lya/$r$ ratio and of the \lyba\ ratio 
for Ne lines (O lines show analogous behaviour).
The \lya/$r$ is much more temperature-sensitive than \lyba, and the
relatively small deviations from the optically-thin case that might be
expected in stellar coronae can only incur small errors in temperature
that will have a negligible impact on the predicted \lyba\ ratio.

\clearpage
\clearpage

The accurate temperature determination from the \lya/$r$ ratio 
strictly holds only for isothermal plasma, whereas stellar coronae 
are characterised by a thermal distribution of the plasma, so that 
this diagnostic would give us a temperature weighted by the emission 
measure distribution (DEM).
By computing line ratios for non-isothermal plasma models, and 
interpreting as if isothermal, we can estimate the errors incurred
by an isothermal assumption.
We considered two different sets of DEMs: those derived from actual 
observations of some of the stars in our sample, and simple DEM models
in which the emission measure is proportional to $T^{3/2}$ (as 
expected for simple hydrostatic loop models; \citealt{RTV}), or 
proportional to $T^{5/2}$ (as observed in some stars such as 31~Com, 
\citealt{Scelsi04}, and reproduced by some hydrodynamic loop models, 
\citealt{Testa05}).  For these models we used peak temperatures 
varying from $10^{6.5}$~K to $10^{7.3}$~K.
Observed DEMs were culled from the literature for the following
sources: AB~Dor \citep{Sanz03b}, HD~223460 (\citealt{Testa07}), 
31~Com, $\beta$~Cet, $\mu$~Vel 
\citep{GarciaA06}, ER~Vul, TZ~CrB, $\xi$~UMa \citep{Sanz03a}, 44~Boo
\citep{Brickhouse98}, UX~Ari \citep{Sanz02}, II~Peg 
\citep{Huenemoerder01}, $\lambda$~And \citep{Sanz02}, AR~Lac 
\citep{Huenemoerder03}, HR~1099 \citep{Drake01}.

\begin{figure}[!ht]
\centerline{\includegraphics[width=8cm,angle=90]{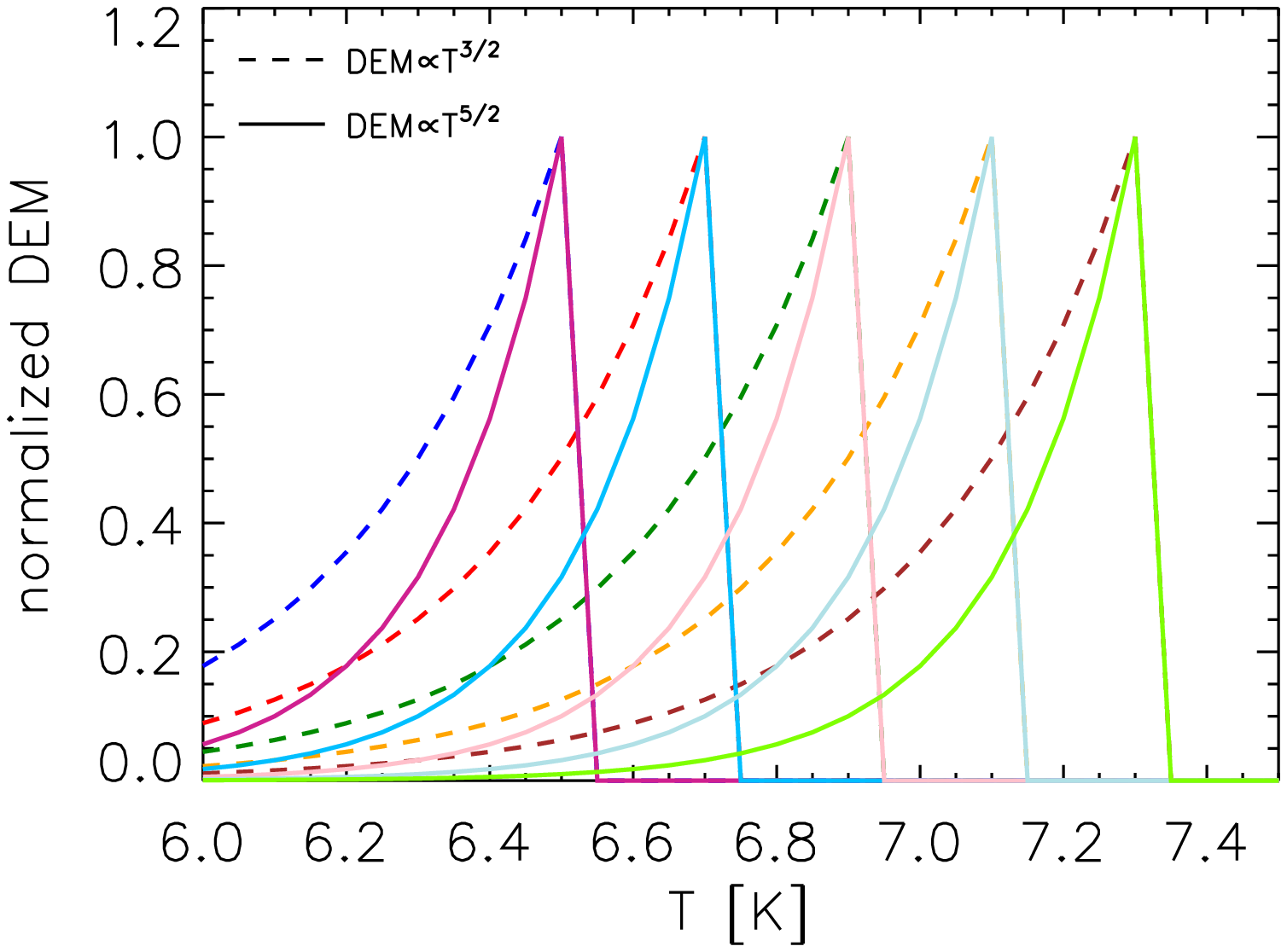}\hspace{-0.8cm}
	\includegraphics[width=8cm,angle=90]{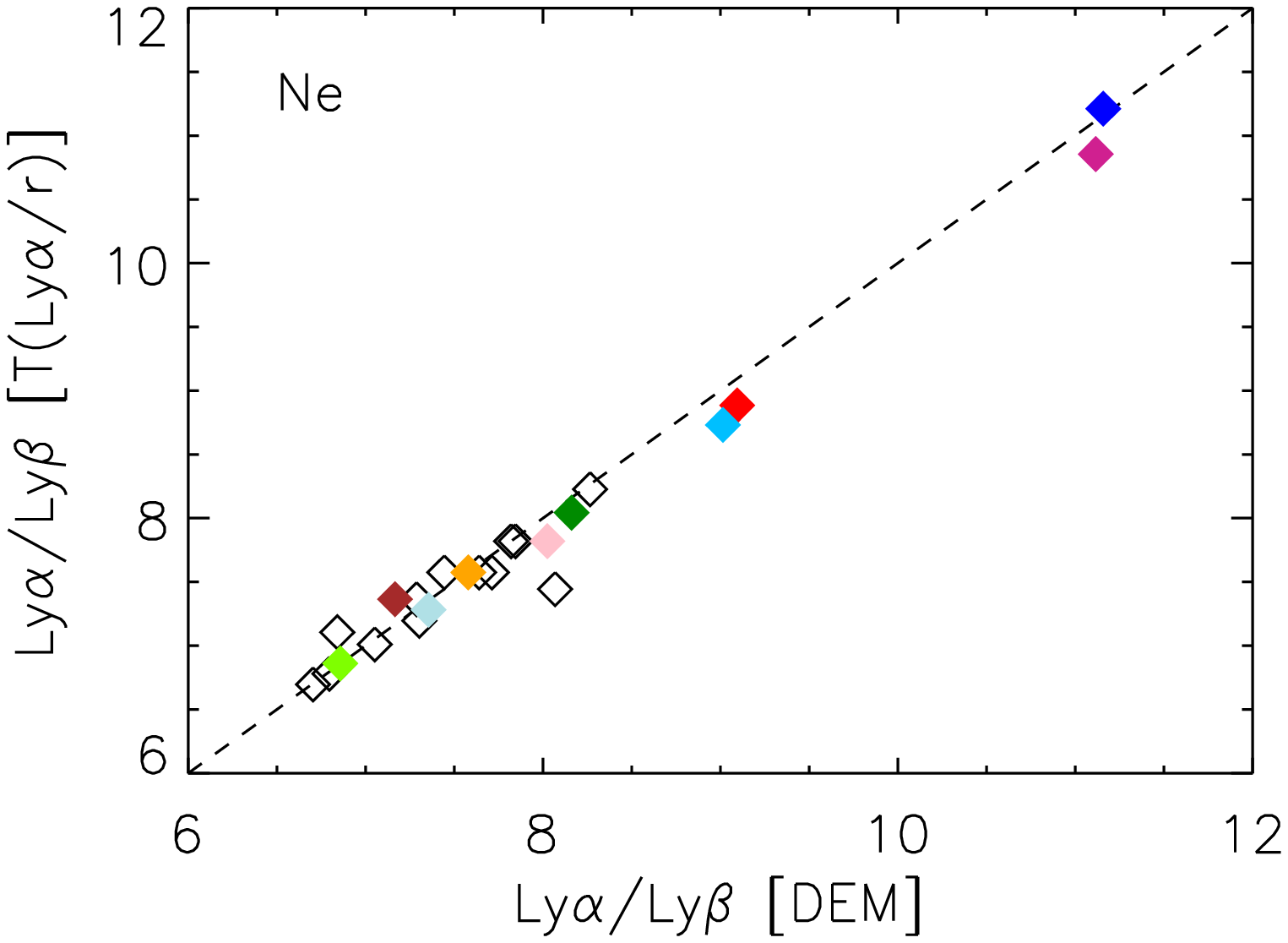}}\vspace{-0.5cm}
\caption{{\em Left panel:} DEM models (DEM $\propto T^{3/2}, T^{5/2}$) 
	used to test the validity of the isothermal approximation.
	{\em Right panel:} The \nex\ \lyab\ ratios obtained for model
	(left panel) and observed DEMs (see text; black empty symbols) 
	compared with the \lyab\ ratios expected for isothermal 
	temperatures diagnosed from the DEM Ne \lya/$r$ ratios. 
	\label{fig:DEMvsIsot}}
\end{figure}

Given the line ratio computed for each DEM model, we inverted the
isothermal relation to obtain $T($\lya/$r)$.  Using that $T$, we
then obtained the theoretical isothermal \lyab\ ratio 
(\lyab\ [$T($\lya/$r)$]) and compared to the synthetic ratio 
(\lyab\ [DEM]). 
We find that the isothermal assumption is a good predictor of the 
ratio for both theoretical DEMs and for representative stellar models: 
Figure~\ref{fig:DEMvsIsot} shows that the isothermal and DEM ratios
are in very good agreement, with differences between the two 
generally amounting to a few percent.  We conclude that the \lya/$r$
temperatures are quite adequate to assess the appropriate expected
\lyab\ ratio, even if the real plasma temperature distributions of 
the coronae in our study are far from isothermality.

\clearpage
\clearpage

\subsection{Deblending of \nex\ \lya\ and \oviii\ \lyb}

\nex\ \lya\ and \oviii\ \lyb\ lines are affected by blending of iron
lines unresolved at the \hetgs\ resolution level.  Specifically,
significant blending is expected for \oviii\ \lyb\ by an \fexviii\ line 
($2s^22p^4(^3P)3s~^2P_{3/2} - 2s^22p^5~^2P_{3/2}$, $\lambda=16.004$\AA), 
and, to a lesser extent, for the \nex\ \lya\ line by a nearby \fexvii\ 
transition 
($2s^22p^5(^2P)4d~^1P_{1} - 2s^22p^6~^1S_{0}$, $\lambda=12.124$\AA).

\paragraph{\oviii\ \lyb~---}
In order to estimate the often significant contribution of \fexviii\
to the \oviii\ \lyb\ spectral feature, we used the same method
described in \ta, which was also subsequently adopted by 
\citet{Ness05} in an analysis of spectra of the classical T~Tauri star
TW~Hya.  We estimated the intensity of the 16.004\AA\ \fexviii\ line
by scaling the observed intensity of the slightly stronger 
neighbouring unblended \fexviii\ 16.071~\AA\ 
($2s^2 2p^4(^3P)3s~^4P_{5/2} - 2s^2 2p^5~^2P_{3/2}$)
transition.  \citet{Gu03} has recently pointed out the similar
behaviour of Fe L-shell lines originating from $3s$ and $3p$ upper
levels in respect to the indirect excitation processes of radiative 
recombination, dielectronic recombination, and resonance excitation.  
The 16.004~\AA\ and 16.071~\AA\ transitions originate from similar 
$3s$ upper levels and their ratio should therefore not deviate greatly
from current theoretical predictions.  The APED (v.1.3.1) database
\citep{Smith01} lists their theoretical ratio as 0.76 at the
temperature of the \fexviii\ population peak ($\sim 6.8$~MK), with a
decrease of only a few percent up to 10~MK or so; this range covers 
the expected formation temperatures of \fexviii\ in our target stars.

\begin{figure}[!ht]
\centerline{\includegraphics[width=8.5cm]{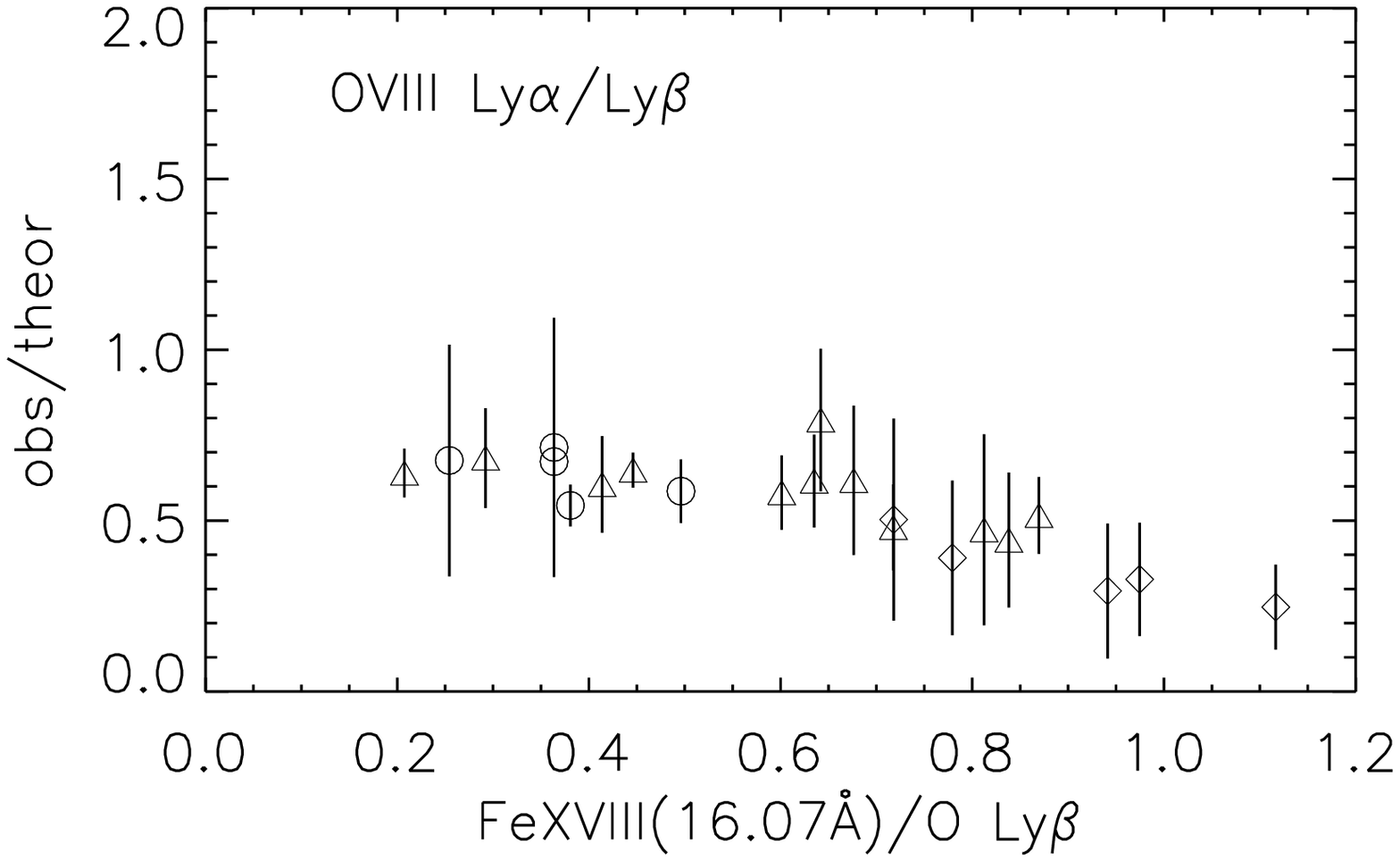}\hspace{-0.7cm}
	\includegraphics[width=8.5cm]{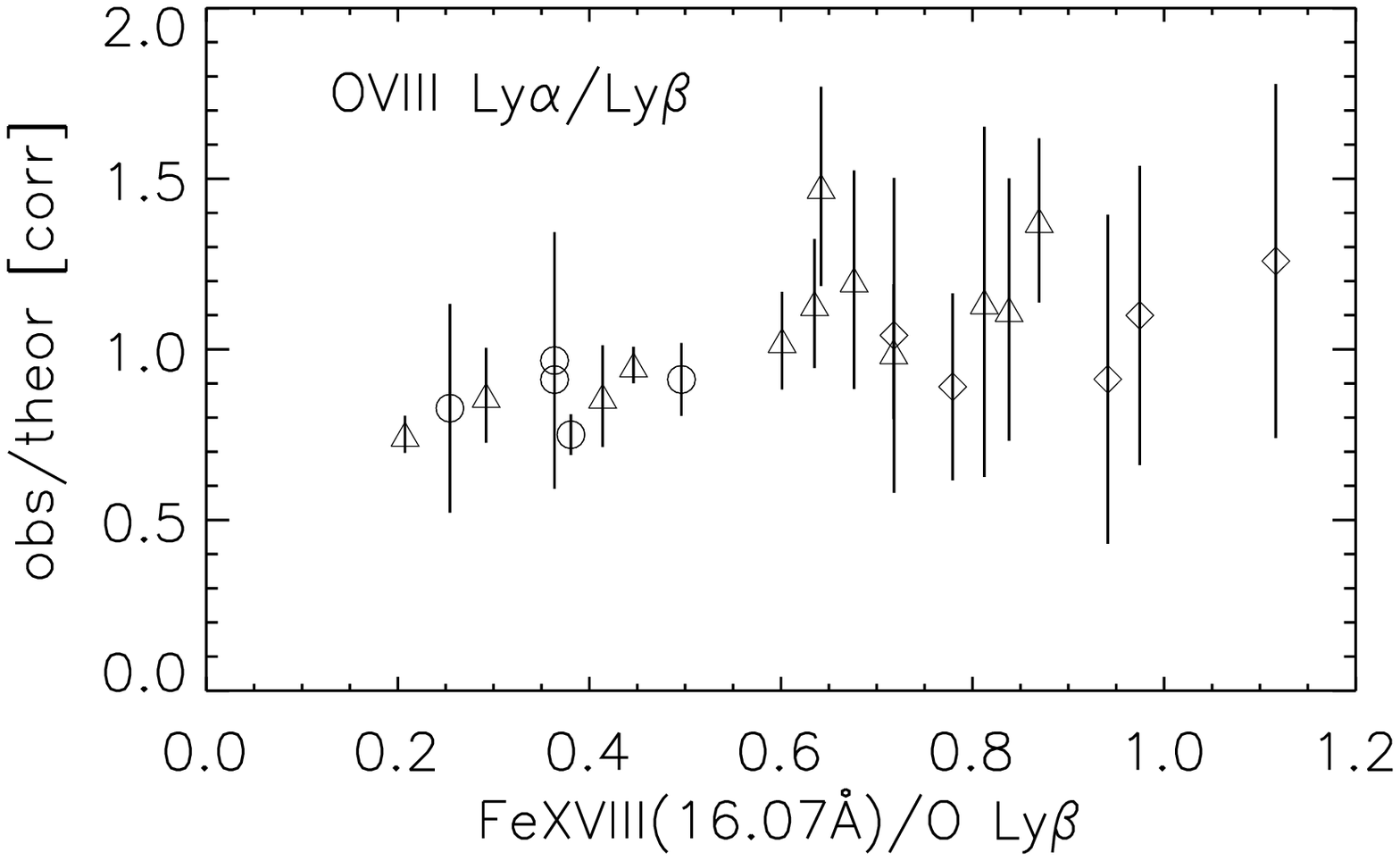}}\vspace{-0.5cm}
\caption{\oviii\ \lyab\ ratios (relative to the theoretical ratio) 
	before ({\em left}) and after ({\em right}) the correction for
	the contamination of the \fexviii\ line blending with the O 
	\lyb\ line, plotted vs.\ the relative intensity of the 
	\fexviii\ line at $\sim$16.07~\AA\ and the O \lyb\ before the 
	deblending process. The effect of the blending \fexviii\ line 
	is clear from the {\em left} plot. The right plot shows the 
	effectiveness of the deblending procedure which eliminates the 
	strong correlation between the departure from the theoretical 
	\lyab\ value and the \fexviii/O\lyb\ ratio before the 
	deblending.
	\label{fig:corr}}
\end{figure}

We investigated this scaling factor empirically by examining the
departure of the resulting deblended \lyab\ ratios from the 
theoretical value as a function of the ratio of the observed
\fexviii\ 16.071\AA\ to \lyb\ line strengths.  If the deblending
scaling factor is correct, there should be no residual slope in the
resulting data points; a positive slope would instead indicate an
over-correction for the blend and a negative slope an
under-correction.  We found that a slope of zero was obtained for a
scaling factor of $\sim 0.70$---in good agreement with the APED
value within expected uncertainties.  This is illustrated in
Figure~\ref{fig:corr}.  

\clearpage
\clearpage

\paragraph{\nex\ \lya~---}
In analogy with the procedure used for the \oviii\ \lyb\ we searched
for isolated and strong \fexvii\ lines to estimate the amount of
blending from \fexvii\ $\lambda=12.124$\AA\ in the \nex\ \lya\ line.
A potentially good candidate would be the nearby 12.266\AA\ \fexvii\
line ($2s^22p^5(^2P)4d~^3D_{1} - 2s^22p^6~^1S_{0}$), which has a
similar intensity and behavior as a function of temperature; however
this transition is rather weak in most of the observed spectra, and
can be affected itself by blending with a close \fexxi\ transition
(12.284\AA) barely resolved by \hetg.

We therefore chose to use a larger set of \fexvii\ lines including
stronger transitions.  The intensity of the \fexvii\ line blending
with the \nex\ \lya\ is estimated by scaling the observed intensity of
other \fexvii\ lines (12.266, 15.014, 15.261, 16.780, 17.051,
17.096\AA) by the ratio expected from \citet{LandiGu06} (see also
\citealt{GuFe17,chianti06}), assuming $\log T = 6.8$.  These
calculations include indirect processes involving the neighboring
charge states, and are quite successful at reproducing the relative
intensity of these \fexvii\ lines.  Previous calculations (including
those adopted in APED v.1.3.1) fail here, with, e.g., the ratio of the
15\AA\ line to the 17\AA\ lines being overestimated by about a factor
2.  The detailed comparison of the observed \fexvii\ line ratios with
the predictions of different theoretical calculations are addressed in
a paper in preparation.

The measured line fluxes after application of deblending corrections,
together with the statistical errors, are listed in
Table~\ref{tab:linefluxes}.

\section{Results}
\label{ss:results}

\begin{figure*}[!ht]
\centerline{\includegraphics[width=10cm,height=6.5cm]{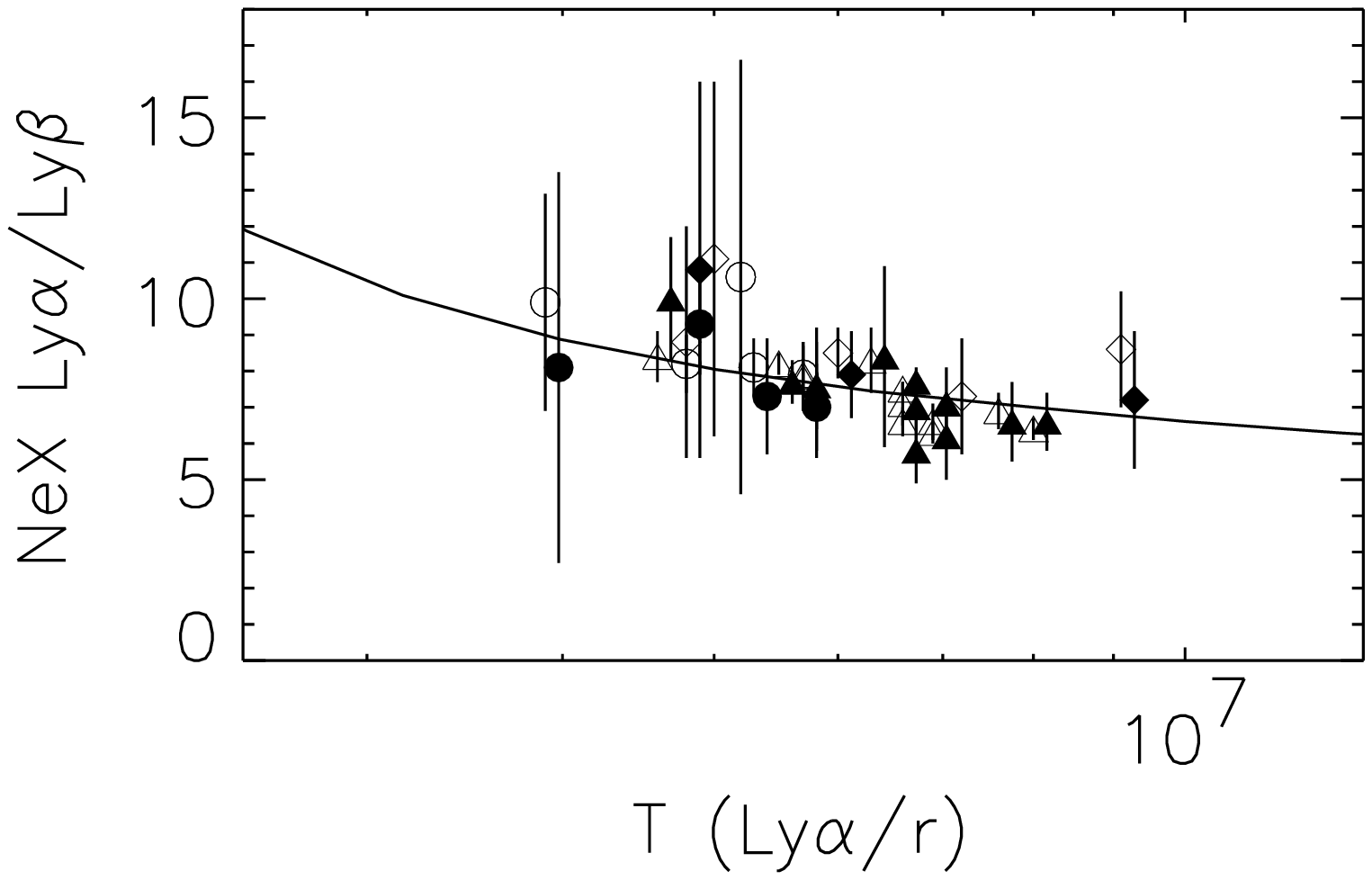}\hspace{-1.5cm}
       \includegraphics[width=10cm,height=6.5cm]{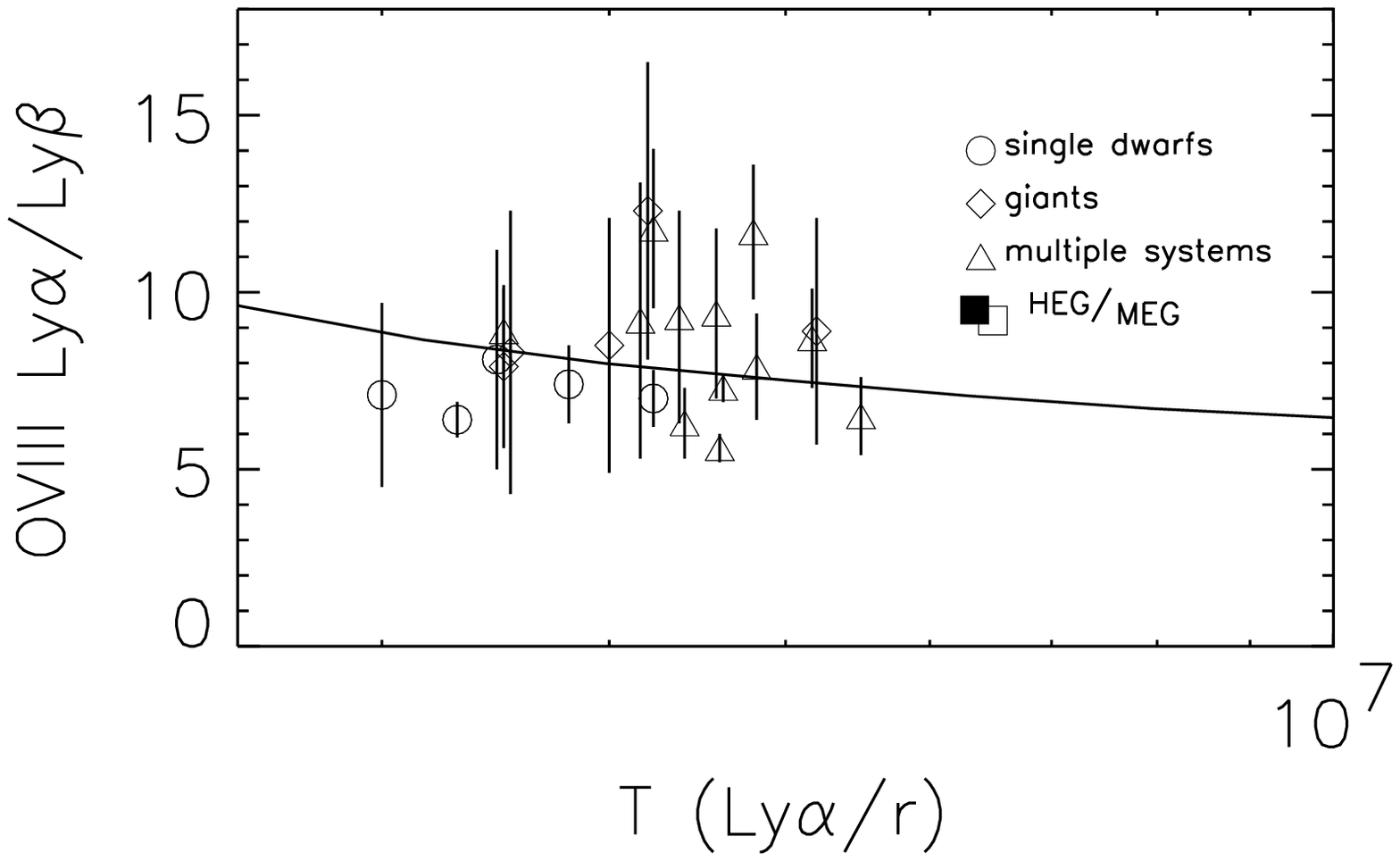}}
\caption{\lyab, both from {\sc heg} (filled symbols) and from 
	{\sc meg} (empty symbols) vs.\ temperature, as derived from 
	the ratio of the \lya\ and the He-like resonance line. 
	The solid line mark the theoretical ratio from APED, expected 
	for isothermal plasma at the corresponding temperature. As 
	indicated on the right plot, different symbols are used for 
	different classes of sources. \lyab\ ratios from {\sc heg} are
	shifted in T by 2\% to separate them from the {\sc meg} 
	measurements.
	\label{fig:allresLyab_temp}}
\end{figure*}

The observed \lyab\ ratios and the temperature derived
from the \lya/$r$ ratios are listed in Table~\ref{tab:lyabT} and
illustrated in Figure~\ref{fig:allresLyab_temp}.
Measurements obtained from the {\sc meg} spectra generally result in
smaller statistical errors due to the higher S/N; however, we find
a general agreement of {\sc heg} and {\sc meg} results\footnote{The 
issue of some systematic discrepancies between {\sc heg} and {\sc meg} 
measurements has been addressed briefly in \tb, and it is thoroughly 
discussed in {http://space.mit.edu/ASC/calib/heg\_meg/}, where 
systematic discrepancies of 8-10\% are found in the 10-12\AA\ range 
(\meg\ fluxes being lower than corresponding \heg\ fluxes at the same 
wavelength) therefore not significantly affecting the \lya/\lyb\ ratios.
We note that our measured \heg\ fluxes are in agreement with \meg\ 
measurements within $2 \sigma$ for the large majority of cases;
the only significant exceptions are the \nex\ \lya\ of TZ~CrB and of 
HR~1099 ($3, 5 \sigma$ respectively) whose \heg\ fluxes are $\sim 10$\% 
larger than the corresponding \meg\ fluxes in agreement with the 
comparisons presented in {http://space.mit.edu/ASC/calib/heg\_meg/}.
}.
The main differences of the present work with respect to the analysis 
of \ta, where results were presented for 4 sources (II~Peg, 
IM~Peg, HR~1099, and AR~Lac), are that the line fluxes have been 
remeasured in the reprocessed data, \nex\ \lya\ fluxes have been 
corrected for blending, and the deblending procedure for \oviii\ \lyb\ 
has been slightly refined (see previous section). Also, IM~Peg
fluxes listed here refer to the total ($\sim 192$~ks) \hetgs\ spectrum 
including ObsID 2530-2534 not yet publicly available when we carried 
out the analysis presented in \ta\ (where we used the first three 
25~ks pointings; see \S\ref{sec:IMPeg} for a detailed discussion).

\clearpage
\begin{deluxetable}{lrrrrrrrrrrrrrrrrrrr}
\tablecolumns{18} 
\tabletypesize{\tiny}
\tablecaption{Measured line fluxes (in $10^{-6}$~photons~cm$^{-2}$~s$^{-1}$).
	\label{tab:linefluxes}}
\tablewidth{0pt}
\tablehead{
 \colhead{Source} & \multicolumn{7}{c}{Ne} & \colhead{} &  \multicolumn{5}{c}{O} & \colhead{} &  \multicolumn{3}{c}{Fe} \\
 \cline{2-8} \cline{10-14} \cline{16-18} \\[-0.15cm]
 \colhead{} & \multicolumn{2}{c}{Ly$\alpha$ \tablenotemark{a}} & \colhead{} & 
 \multicolumn{2}{c}{Ly$\beta$} & \colhead{} & \colhead{$r$} & \colhead{} & \colhead{Ly$\alpha$} 
 & \colhead{} & \colhead{Ly$\beta$ \tablenotemark{b}} & \colhead{} & \colhead{$r$} & \colhead{} 
 & \colhead{\fexvii \tablenotemark{c}} & \colhead{} & \colhead{\fexviii} \\
 \colhead{} & \multicolumn{2}{c}{12.132\AA} & \colhead{} & \multicolumn{2}{c}{10.239\AA} & 
 \colhead{} & \colhead{13.447\AA} & \colhead{} & \colhead{18.967\AA} & \colhead{} & 
 \colhead{16.006\AA} & \colhead{} & \colhead{21.602\AA} & \colhead{} & \colhead{12.124\AA} & 
 \colhead{} & \colhead{16.071\AA} \\
 \cline{2-3} \cline{5-6} \cline{8-8} \cline{10-10} \cline{12-12} \cline{14-14} 
 \cline{16-16} \cline{18-18}\\[-0.15cm]
 \colhead{}    & \colhead{HEG} & \colhead{MEG} & \colhead{}    & \colhead{HEG} & 
 \colhead{MEG} & \colhead{}    & \colhead{MEG} & \colhead{}    & \colhead{MEG} & 
 \colhead{}    & \colhead{MEG} & \colhead{}    & \colhead{MEG} & \colhead{}    & 
 \colhead{MEG} & \colhead{} & \colhead{MEG} 
}
\startdata 
 AU~Mic 	& $374 \pm 50$  & $356 \pm 15$  & & $51 \pm 10$  & $44 \pm 4$     & & $208 \pm 15$   & & 
 		  $990 \pm 40$  & & $134 \pm 20$ & & $243 \pm 40$ & & $21  \pm 5$ & & $67  \pm 11$ \\
 Prox~Cen 	& -             & $ 69 \pm 20$  & & -            & $6.5 \pm 2.9$  & & $47 \pm 9$     & & 
 		  $306 \pm 28$  & & $38 \pm 14$  & & $100 \pm 70$ & & $6   \pm 2$ & & $19  \pm 13$ \\
 EV~Lac 	& $278 \pm 22$  & $262 \pm 10$  & & $30 \pm 6$   & $32 \pm 3$     & & $209 \pm 8$    & & 
 		  $875 \pm 19$  & & $137 \pm 11$ & & $330 \pm 30$ & & $24  \pm 5$ & & $73  \pm  9$ \\
 AB~Dor 	& $668 \pm 50$  & $682 \pm 24$  & & $96 \pm 16$  & $86 \pm 9$     & & $332 \pm 27$   & & 
 		  $1400 \pm 50$ & & $200 \pm 22$ & & $280 \pm 50$ & & $41  \pm 9$ & & $158 \pm 16$ \\
 TW~Hya 	& $97 \pm 19$   & $95 \pm 8$    & & $12 \pm 8$   & $9.6 \pm 2.8$  & & $155 \pm 14$   & & 
 		  $265 \pm 30$  & & $38 \pm 12$  & & $114 \pm 40$ & & $5.9 \pm 2$ & & $11  \pm  8$ \\
 HD~223460 	& $158 \pm 20$  & $137 \pm 11$  & & $22 \pm 6$   & $16.0 \pm 2.7$ & & $20 \pm 7$     & & 
 		  $229 \pm 22$  & & $27 \pm 11$  & & $48 \pm 24$  & & $10  \pm 2$ & & $42  \pm 10$ \\
 HD~111812 	&               & $58 \pm 5$    & & -            & $7.9 \pm 1.6$  & & $14 \pm 3$     & & 
 		  $124 \pm 16$  & & $15 \pm 7$   & & $40 \pm 30$  & & $8.8 \pm 2$ & & $48  \pm  5$ \\
 $\beta$~Cet 	& $419 \pm 24$  & $408 \pm 10$  & & $53 \pm 10$  & $48 \pm 4$     & & $146 \pm 10$   & & 
 		  $764 \pm 29$  & & $62 \pm 23$  & & $140 \pm 40$ & & $116\pm 21$ & & $426 \pm 21$ \\
 HD~45348 	& -             & $54 \pm 4$    & & -            & $4.7 \pm 2.3$  & & $35 \pm 7$     & & 
 		  $127 \pm 22$  & & $16 \pm 8$   & & $40 \pm 30$  & & $11.6\pm 3$ & & $ 30 \pm  5$ \\
 $\mu$~Vel 	& $173 \pm 21$  & $159 \pm 15$  & & $16 \pm 13$  & $18 \pm 7$     & & $95 \pm 7$     & & 
 		  $400 \pm 30$  & & $45 \pm 16$  & & $46 \pm 27$  & & $62 \pm 12$ & & $163 \pm 10$ \\
 Algol 		& $898 \pm 50$  & $832 \pm 40$  & & $136 \pm 23$ & $120 \pm 8$    & & $196 \pm 20$   & & 
 		  $1318 \pm 50$ & & $166 \pm 30$ & & $190 \pm 60$ & & $63 \pm 14$ & & $258 \pm 23$ \\
 ER~Vul 	& $218 \pm 22$  & $201 \pm 12$  & & $26 \pm 8$   & $24.3 \pm 2.8$ & & $74 \pm 7$     & & 
 		  $362 \pm 29$  & & $39 \pm 12$  & & $66 \pm 40$  & & $24  \pm 5$ & & $ 88 \pm  9$ \\
 44~Boo 	& $572 \pm 60$  & $537 \pm 24$  & & $75 \pm 11$  & $68 \pm 8$     & & $254 \pm 20$   & & 
 		  $1450 \pm 60$ & & $122 \pm 22$ & & $266 \pm 60$ & & $53 \pm 11$ & & $151 \pm 16$ \\
 TZ~CrB 	& $1126 \pm 30$ & $1043 \pm 24$ & & $145 \pm 14$ & $127 \pm 6$    & & $516 \pm 20$   & & 
 		  $2110 \pm 50$ & & $178 \pm 26$ & & $300 \pm 50$ & & $131\pm 25$ & & $445 \pm 20$ \\
 UX~Ari 	& $730 \pm 40$  & $767 \pm 20$  & & $126 \pm 17$ & $107 \pm 8$    & & $245 \pm 18$   & & 
 		  $926 \pm 50$  & & $147 \pm 21$ & & $158 \pm 40$ & & $14.5\pm 3$ & & $ 55 \pm 13$ \\
 $\xi$~UMa 	& $419 \pm 27$  & $394 \pm 13$  & & $42 \pm 8$   & $47 \pm 4$     & & $309 \pm 16$   & & 
 		  $1510 \pm 60$ & & $169 \pm 22$ & & $496 \pm 60$ & & $83 \pm 17$ & & $185 \pm 15$ \\
 II~Peg 	& $1300 \pm 70$ & $1200 \pm 30$ & & $183 \pm 24$ & $191 \pm 10$   & & $356 \pm 20$   & & 
 		  $1950 \pm 70$ & & $350 \pm 30$ & & $250 \pm 50$ & & $18 \pm  4$ & & $ 86 \pm 16$ \\
 $\lambda$~And 	& $557 \pm 30$  & $556 \pm 14$  & & $80 \pm 11$  & $84 \pm 5$     & & $189 \pm 14$   & & 
 		  $984 \pm 40$  & & $113 \pm 17$ & & $110 \pm 30$ & & $24.6\pm 6$ & & $137 \pm 12$ \\
 TY~Pyx 	& $341 \pm 29$  & $312 \pm 6$   & & $46 \pm 12$  & $40 \pm 5$     & & $142 \pm 15$   & & 
 		  $468 \pm 50$  & & $51 \pm 21$  & & $96 \pm 50$  & & $25  \pm 5$ & & $106 \pm 17$ \\
 AR~Lac 	& $648 \pm 40$  & $621 \pm 17$  & & $104 \pm 15$ & $94 \pm 6$     & & $182 \pm 20$   & & 
 		  $888 \pm 50$  & & $94 \pm 23$  & & $130 \pm 50$ & & $34  \pm 7$ & & $129 \pm 14$ \\
 HR~1099 	& $1960 \pm 40$ & $1730 \pm 18$ & & $255 \pm 14$ & $230 \pm 7$    & & $566 \pm 18$   & & 
 		  $2680 \pm 50$ & & $367 \pm 21$ & & $410 \pm 40$ & & $58 \pm 12$ & & $246 \pm 14$ \\
 IM~Peg 	& $408 \pm 25$  & $383 \pm 10$  & & $62 \pm 7$   & $60 \pm 3$     & & $79 \pm 8$     & & 
 		  $490 \pm 30$  & & $75 \pm 12$  & & $50 \pm 20$  & & $10.6\pm 2.5$ & & $ 45 \pm  8$ 
 \enddata
\tablenotetext{a}{Ne Ly$\alpha$ fluxes after the deblending with the \fexvii\ line at 12.124\AA.}
\tablenotetext{b}{O Ly$\beta$ fluxes after the deblending with the \fexviii\ line at 16.004\AA.}
\tablenotetext{c}{\fexvii\ line (at 12.124\AA) flux estimated by scaling the measured fluxes of the other \fexvii\ lines.}
\end{deluxetable}

\clearpage
\clearpage
\begin{deluxetable}{lrrrrrrrrrr}
\tablecolumns{11} 
\tabletypesize{\footnotesize}
\tablecaption{\lyab\ photon ratios and plasma temperature, with 1$\sigma$ errors.
	\label{tab:lyabT}}
\tablewidth{0pt}
\tablehead{
 \colhead{Source} & \multicolumn{6}{c}{Ne} & \colhead{} &  \multicolumn{3}{c}{O} \\
 \cline{2-7} \cline{9-11} \\[-0.15cm]
 \colhead{} & \colhead{$T$ \tablenotemark{a}} & \multicolumn{2}{c}{\lyab} & 
 \colhead{} & \multicolumn{2}{c}{$\Delta [\sigma]$ \tablenotemark{b}} & \colhead{} &  
 \colhead{$T$ \tablenotemark{a}} & \colhead{\lyab} & \colhead{$\Delta [\sigma]$ \tablenotemark{b}} \\
 \cline{3-4} \cline{6-7} \\[-0.15cm]
 \colhead{}    & \colhead{[$10^6$~K]} & \colhead{HEG} & \colhead{MEG} & \colhead{} & 
 \colhead{HEG} & \colhead{MEG} & \colhead{} & \colhead{[$10^6$~K]} & \colhead{MEG} & \colhead{MEG}
}
\startdata 
 AU~Mic 	& $5.3^{+0.2}_{-0.1}$  & $7.3 \pm 1.6$  & $8.1 \pm 0.8$  & & -0.4 &  0.1 & & $3.8^{+0.2}_{-0.2}$ & $7.4 \pm 1.1$   & -0.7 \\
 Prox~Cen 	& $5.2^{+0.5}_{-0.7}$  & ...		& $10.6 \pm 6$   & & ...  &  0.4 & & $3.5^{+0.7}_{-0.9}$ & $8.1 \pm 3.1$   & -0.1 \\
 EV~Lac 	& $4.8^{+0.1}_{-0.1}$  & $9.3 \pm 1.9$  & $8.2 \pm 0.8$  & &  0.5 & -0.1 & & $3.3^{+0.1}_{-0.2}$ & $6.4 \pm 0.5$   & -4   \\
 AB~Dor 	& $5.7^{+0.1}_{-0.2}$  & $7.0 \pm 1.3$  & $7.9 \pm 0.9$  & & -0.6 &  0.3 & & $4.2^{+0.2}_{-0.4}$ & $7.0 \pm 0.8$   & -0.9 \\
 TW~Hya 	& $3.9^{+0.1}_{-0.1}$  & $8.1 \pm 5.6$  & $9.9 \pm 3.0$  & & -0.1 &  0.3 & & $3.0^{+0.3}_{-0.4}$ & $7.1 \pm 2.6$   & -0.7 \\
 HD~223460 	& $9.1^{+1.3}_{-1.5}$  & $7.2 \pm 2.1$  & $8.6 \pm 1.6$  & &  0.3 &  1.2 & & $4.0^{+0.8}_{-0.8}$ & $8.5 \pm 3.6$   &  0.2 \\
 HD~111812 	& $7.2^{+0.7}_{-0.6}$  & ...		& $7.3 \pm 1.6$  & & ...  &  0.1 & & $3.5^{+0.7}_{-1.2}$ & $8.3 \pm 4.0$   &  0   \\
 $\beta$~Cet 	& $6.0^{+0.2}_{-0.1}$  & $7.9 \pm 1.5$  & $8.5 \pm 0.7$  & &  0.2 &  1.2 & & $4.2^{+0.6}_{-0.4}$ & $12.3 \pm 4.2$  &  1.0 \\
 HD~45348 	& $5.0^{+0.2}_{-0.5}$  & ...		& $11.2 \pm 5.4$ & & ...  &  0.6 & & $3.5^{+0.9}_{-1.3}$ & $7.9 \pm 2.3$   & -0.2 \\
 $\mu$~Vel 	& $4.8^{+0.2}_{-0.2}$  & $10.8 \pm 6.8$ & $8.8 \pm 3.5$  & &  0.4 &  0.1 & & $5.2^{+1.0}_{-1.6}$ & $8.9 \pm 3.2$   &  0.5 \\
 Algol 		& $7.6^{+0.4}_{-0.4}$  & $6.6 \pm 1.2$  & $6.9 \pm 0.6$  & & -0.4 & -0.3 & & $4.8^{+0.3}_{-0.8}$ & $7.9 \pm 1.5$   &  0.2 \\
 ER~Vul 	& $6.3^{+0.1}_{-0.3}$  & $8.4 \pm 2.7$  & $8.3 \pm 1.1$  & &  0.3 &  0.8 & & $4.4^{+0.8}_{-1.4}$ & $9.3 \pm 3.0$   &  0.5 \\
 44~Boo 	& $5.7^{+0.1}_{-0.2}$  & $7.6 \pm 1.4$  & $7.9 \pm 1.0$  & & -0.1 &  0.2 & & $4.2^{+0.4}_{-0.4}$ & $11.9 \pm 2.2$  &  1.9 \\
 TZ~CrB 	& $5.5^{+0.2}_{-0.1}$  & $7.7 \pm 0.8$  & $8.2 \pm 0.4$  & & -0.1 &  0.9 & & $4.8^{+0.2}_{-0.4}$ & $11.9 \pm 1.8$  &  2.2 \\
 UX~Ari 	& $6.6^{+0.3}_{-0.1}$  & $5.8 \pm 0.9$  & $7.1 \pm 0.6$  & & -1.7 & -0.3 & & $4.4^{+0.4}_{-0.4}$ & $6.3 \pm 1.0$   & -1.3 \\
 $\xi$~UMa 	& $4.6^{+0.2}_{-0.1}$  & $10.0 \pm 2.0$ & $8.4 \pm 0.8$  & &  0.8 &  0.0 & & $3.5^{+0.1}_{-0.2}$ & $8.9 \pm 1.2$   &  0.4 \\
 II~Peg 	& $6.9^{+0.3}_{-0.1}$  & $7.1 \pm 1.0$  & $6.3 \pm 0.4$  & & -0.2 & -2.8 & & $4.6^{+0.4}_{-0.4}$ & $5.6 \pm 0.4$   & -5   \\
 $\lambda$~And 	& $6.6^{+0.1}_{-0.3}$  & $7.0 \pm 1.0$  & $6.6 \pm 0.4$  & & -0.4 & -1.8 & & $5.2^{+0.5}_{-0.6}$ & $8.7 \pm 1.4$   &  0.9 \\
 TY~Pyx 	& $5.7^{+0.3}_{-0.2}$  & $7.4 \pm 1.8$  & $7.8 \pm 0.9$  & & -0.1 &  0.1 & & $4.2^{+0.6}_{-1.0}$ & $9.2 \pm 3.9$   &  0.4 \\
 AR~Lac 	& $6.9^{+0.3}_{-0.3}$  & $6.2 \pm 1.1$  & $6.6 \pm 0.6$  & & -1.0 & -1.2 & & $4.6^{+0.6}_{-0.8}$ & $9.4 \pm 2.4$   &  0.8 \\
 HR~1099 	& $6.6^{+0.1}_{-0.1}$  & $7.7 \pm 0.4$  & $7.5 \pm 0.2$  & &  0.7 &  0.5 & & $4.6^{+0.2}_{-0.2}$ & $7.3 \pm 0.4$   & -0.6 \\
 IM~Peg 	& $8.0^{+0.4}_{-0.3}$  & $6.6 \pm 0.8$  & $6.4 \pm 0.3$  & & -0.5 & -1.8   & & $5.5^{+0.8}_{-1.1}$ & $6.5 \pm 1.1$   & -0.8    
 \enddata
\tablenotetext{a}{$T$ derived from the \lya/$r$ line ratio diagnostics.}
\tablenotetext{b}{Discrepancy between \lyab\ measured and theoretical value in units of $\sigma$.}
\end{deluxetable}

\clearpage

The observed \nex\ \lyab\ ratios follow closely the APED theoretical
predictions as a function of temperature.  The values for the \oviii\
ratios (right panel of Fig.~\ref{fig:allresLyab_temp}) show slightly
more scatter, with some departures below theory and some 1-2$\sigma$
excursions above.  We note that an {\em enhancement} of the \lyab\
ratio is possible for the particular geometry in which an emitting
region is optically-thick in some directions but is optically-thin in
the line-of-sight of the observer (see also the radiative transfer
study of \citealt{Kerr04}).  Such a situation is disfavoured by the
isotropic nature of the scattering process in which any line
enhancements through scattering are diluted, roughly speaking, by the
ratio of solid angles of optically-thin to optically-thick 
lines-of-sight.  In the case of loop geometries, such a ratio is much 
larger than unity.  We therefore interpret these small upward 
1-$2~\sigma$ excursions as normal statistical fluctuations.

In the case of the \nex\ lines, the {\sc meg} \lyab\ ratio for II~Peg 
is $\sim 3 \sigma$ lower than the expected value; the ratio derived 
from {\sc heg} is consistent with the {\sc meg} measurement but does 
not depart significantly from theory.  The only sources showing 
\oviii\ \lyab\ significantly ($>3 \sigma$) lower than the theoretical 
ratio are II~Peg and EV~Lac.  We note that the low \lyab\ ratios for 
IM~Peg discussed in \ta\ were found in the spectra obtained in the 
first three pointings while here we analyze the whole set of 
observations; in \S\ref{sec:IMPeg} below we discuss the variability 
observed in \lyab\ ratios for IM~Peg.

One other interesting case is the T~Tauri star, TW~Hya, that shows
very high density, $n_{\mathrm{e}} \gtrsim 10^{12}$~cm$^{-3}$, at the
temperature of \ovii\ lines (e.g., \citealt{Kastner02}).  
In this case, the high densities are generally believed to arise 
from an accretion shock, rather than coronal loops; significant 
scattering effects could in principle provide interesting constraints 
on the dimensions of the shock region. Unfortunately, for this
source the relatively low S/N results in large error bars in the
\lyab\ ratios that do not provide any useful constraints (as was also 
noted in the analysis of \fexvii\ lines by \citealt{Ness05}).

\subsection{IM~Peg}
\label{sec:IMPeg}

The different {\it Chandra} \hetg\ observations of the RS CVn~system
IM~Peg provide an opportunity for studying the optical depth and its
possible variation with time.  This system has been analyzed in great
detail at optical wavelengths by
\citet{Berdyugina99,Berdyugina00}. These works depict a scenario with
stellar spots concentrated mainly close to the polar regions, similar
to those found for many other active systems
\citep[e.g.,][]{Schuessler96,Hatzes96,Vogt99,Hussain02,Berdyugina98}.

\begin{figure}[!ht]
\centerline{\includegraphics[width=13cm]{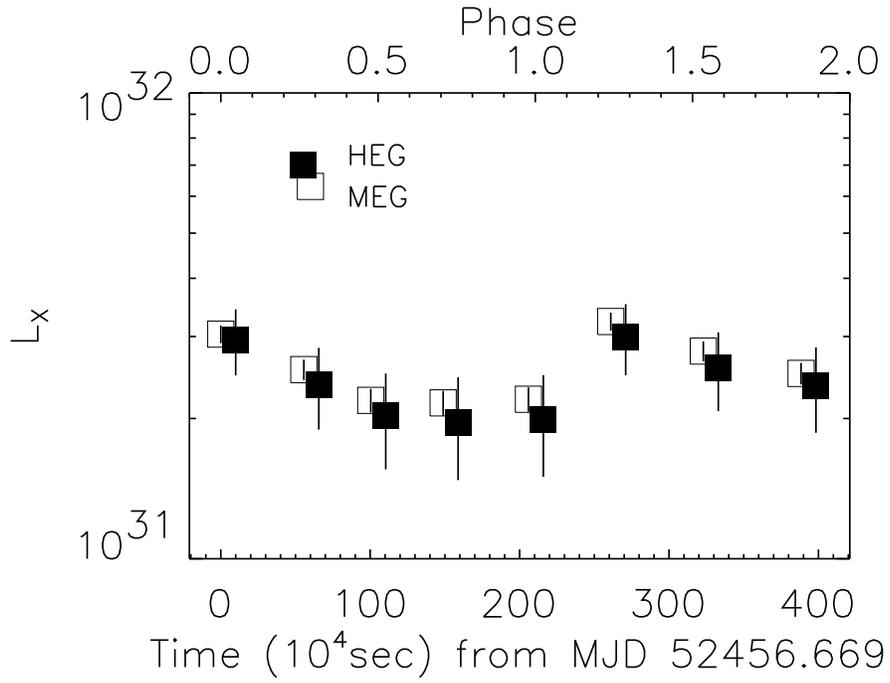}}\vspace{-0.5cm}
\caption{Lightcurve of IM~Peg \cha\ observations; both {\sc heg} 
	({\em filled squares}) and {\sc meg} ({\em empty squares}) 
	\lx\ are presented. {\sc heg} values are shifted by $10^5$~s 
	on the time axis with respect to the corresponding {\sc meg}
	measurements. The phases corresponding to the absolute times 
	are indicated at the top of the plot.
	\label{fig:impeg_lc}}
\end{figure}

\cha\ observed IM~Peg eight times over $\sim 2$ orbital periods 
(see Tab.~\ref{tab_impeg}).  The X-ray lightcurve obtained from this 
sequence is illustrated in Figure~\ref{fig:impeg_lc} and shows 
significant modulation that is possibly related to the orbital and 
rotational period (these being essentially the same; 
$P_{\rm rot}=24.39$~d, $P_{\rm orb}=24.65$~d, \citealt{Strassmeier93}). 
In \ta\ we analyzed the first three 25~ks \hetgs\ observations of 
IM~Peg, publicly available at that time, and we found remarkably small
photon path lengths based on optical depth effects.  
The compact emission regions implied
might be associated with active regions revealed by optical Doppler
imaging studies.  Any significant net line photon loss to resonant
scattering out of the line-of-sight would be expected to be sensitive
to the orientations of the emitting structures.  Modulation of the
\lyab\ ratio with time could provide important new structure and
morphology diagnostics.  Such regions might also be expected to change
on relatively short timescales as a result of flaring activity and
consequent re-alignment of their defining magnetic fields.

\clearpage
\begin{deluxetable}{lrrrr}
\tablecolumns{5} 
\tabletypesize{\footnotesize}
\tablecaption{Ly$\alpha$/Ly$\beta$ photon ratios with 
	1 $\sigma$ errors.
	\label{tabo:ratios_impeg}}
\tablewidth{0pt}
\tablehead{
 \colhead{Obs. ID} & \multicolumn{2}{c}{Ne\,{\sc x}} & 
 \colhead{} & \colhead{O\,{\sc viii}} \\
 \cline{2-3} \cline{5-5} \\[-0.15cm]
 \colhead{} & \colhead{HEG} & \colhead{MEG} & 
 \colhead{} &  \colhead{MEG} 
}
\startdata 
 2527+2528	&  $5.3 \pm 0.8$  &  $ 5.5\pm 0.5$  & &  $5.1 \pm 1.2$ \\
 2529+2530+2531	&  $6.9 \pm 1.7$  &  $ 6.5\pm 0.6$  & &  $7.5 \pm 3.0$ \\
 2532+2533+2534	&  $6.2 \pm 1.2$  &  $ 7.1\pm 0.6$  & &  $7.0 \pm 2.1$ 
 \enddata
\end{deluxetable}

\clearpage

\begin{figure}[!ht]
\centerline{\includegraphics[width=12cm]{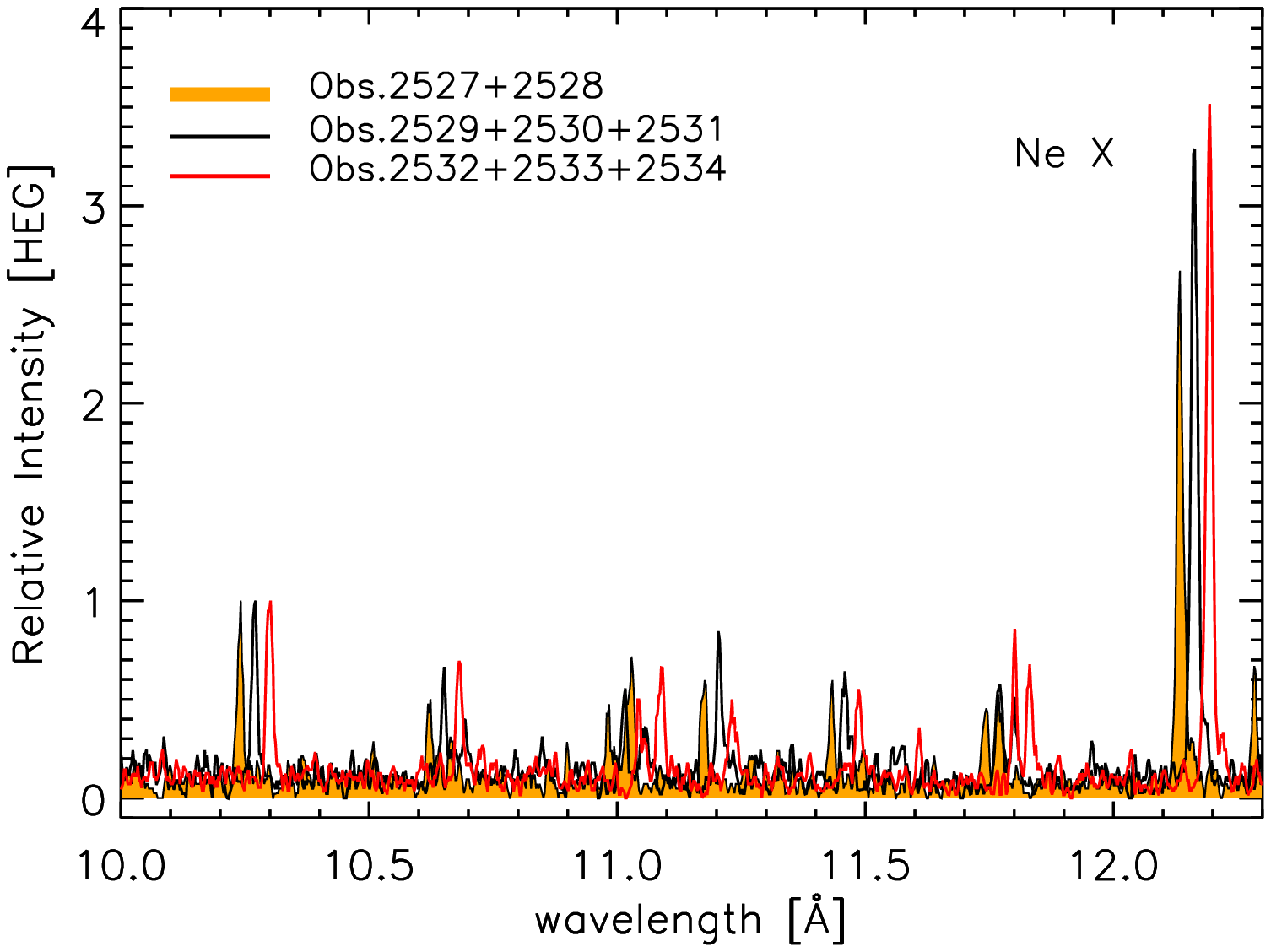}}\vspace{-0.5cm}
\centerline{\includegraphics[width=12cm]{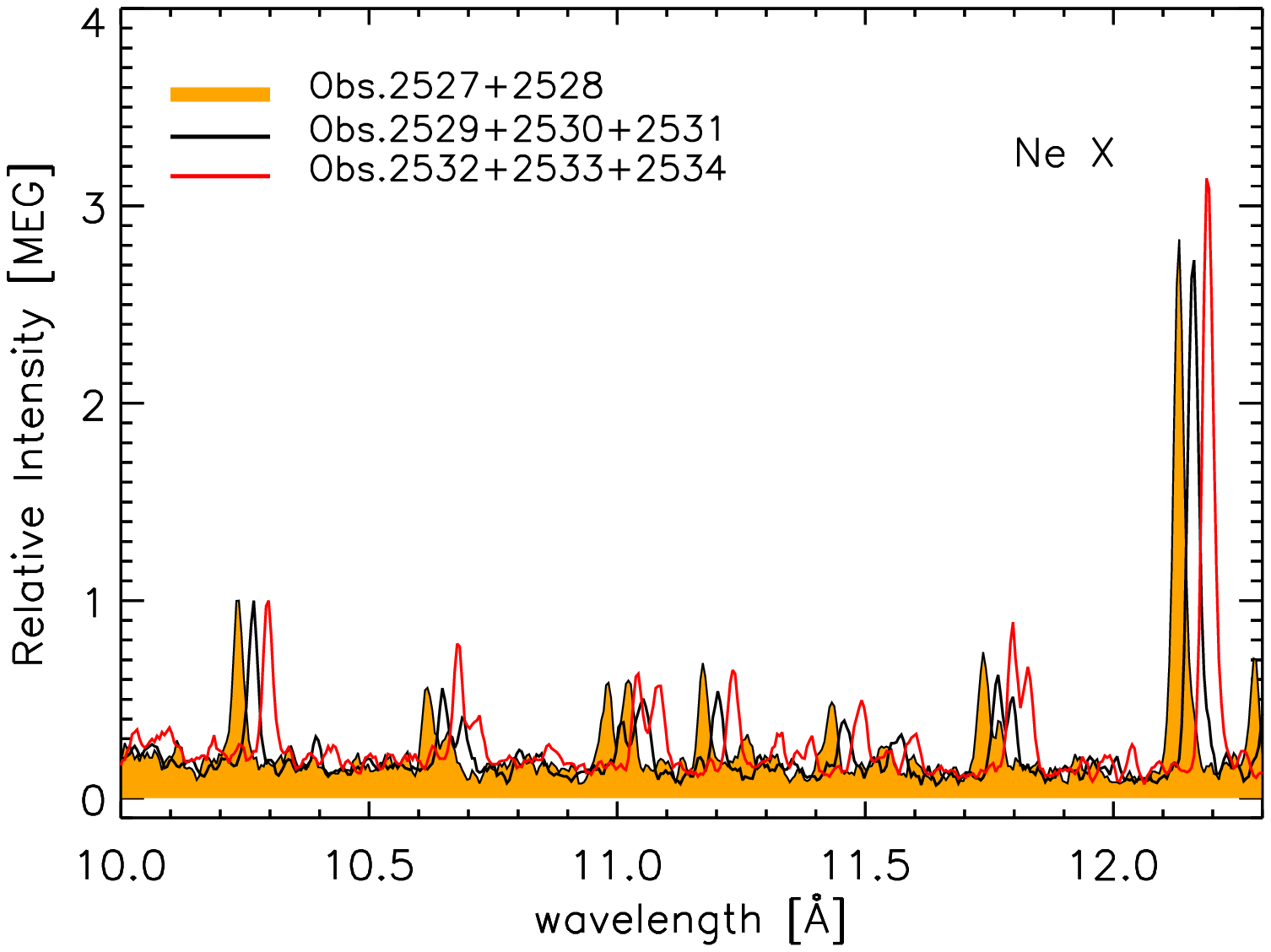}}\vspace{-0.5cm}
\caption{\nex\ Ly$\alpha,\beta$ spectral region for the three chosen
	time segments of the observation of the RS~CVn system IM~Peg, 
	as indicated in the plots.
	Both {\sc heg} ({\em upper panel}) and {\sc meg} ({\em lower 
	panel}) spectra are presented, normalized to the intensity of
	the \lyb\ line at 10.24~\AA. For better readability the 
	spectra	corresponding to the second (2529+2530+2531) and the 
	third (2532+2533+2534) segment are shifted in wavelength with 
	respect to the first (2527+2528) by +0.03~\AA\ and +0.06~\AA\ 
	respectively. Note the pronounced difference in \lya\ and \lyb\ 
	relative intensity in the three spectra especially between the 
	first and the third spectra.
	 \label{fig:impegall_Ne}}
\end{figure}
\clearpage

\begin{figure}[!h]
\centerline{\includegraphics[width=13cm]{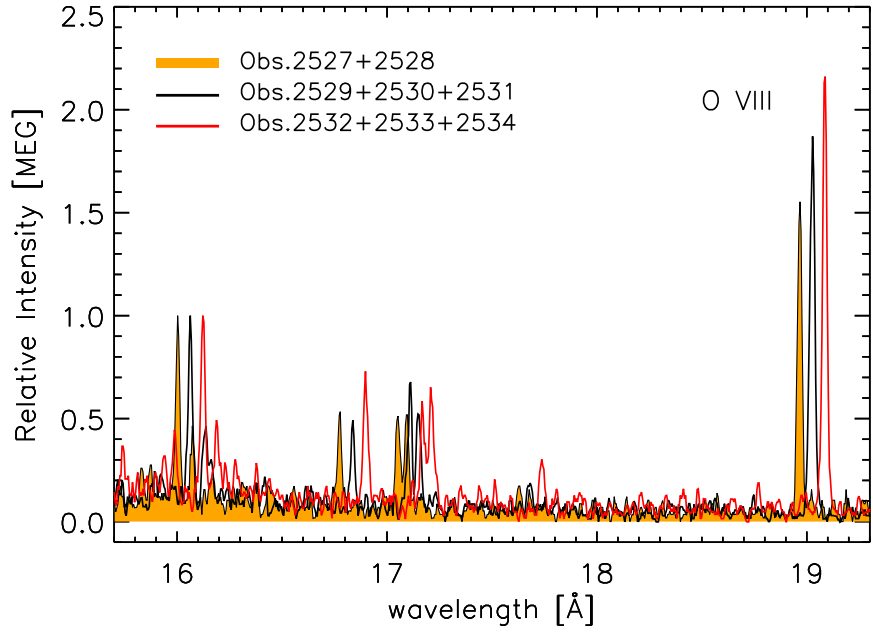}}\vspace{-0.5cm}
\caption{\oviii\ Ly$\alpha,\beta$ spectral region of {\sc meg} spectra 
	of IM~Peg for the three chosen time segment of the observation, 
	in the same format of the plots in Figure~\ref{fig:impegall_Ne}. 
	The shift in wavelength of second and third spectra is of 
	+0.06~\AA\ with respect to the preceding spectrum. A clear 
	trend in the \lya/\lyb\ line ratio is present.
	  \label{fig:impegall_O}}
\end{figure}

Here we present an analysis of the complete dataset, exploring possible 
temporal variability of the optical depth of the coronal plasma.
Unfortunately, the single IM~Peg exposures provided counts sufficient
for reliable intensity measurement only for the strongest lines in the
spectrum, such as \nex\ \lya.  In order to probe secular change in the
\lyab\ ratio, we therefore measured lines from spectra coadded from
contiguous sets of two or three observations that sample different
phases of the stellar period: ObsIDs 2527 and 2528; 2529, 2530, and
2531; and 2532, 2533, and 2534.  The spectra in the Ne and O
Ly$\alpha,\beta$ regions for these three portions of the observation
are shown in Figure~\ref{fig:impegall_Ne}~and~\ref{fig:impegall_O}.
The measured \lyab\ ratios are listed in Table~\ref{tabo:ratios_impeg}.

\clearpage
\clearpage

\begin{figure}[!ht]
\centerline{\includegraphics[width=15cm]{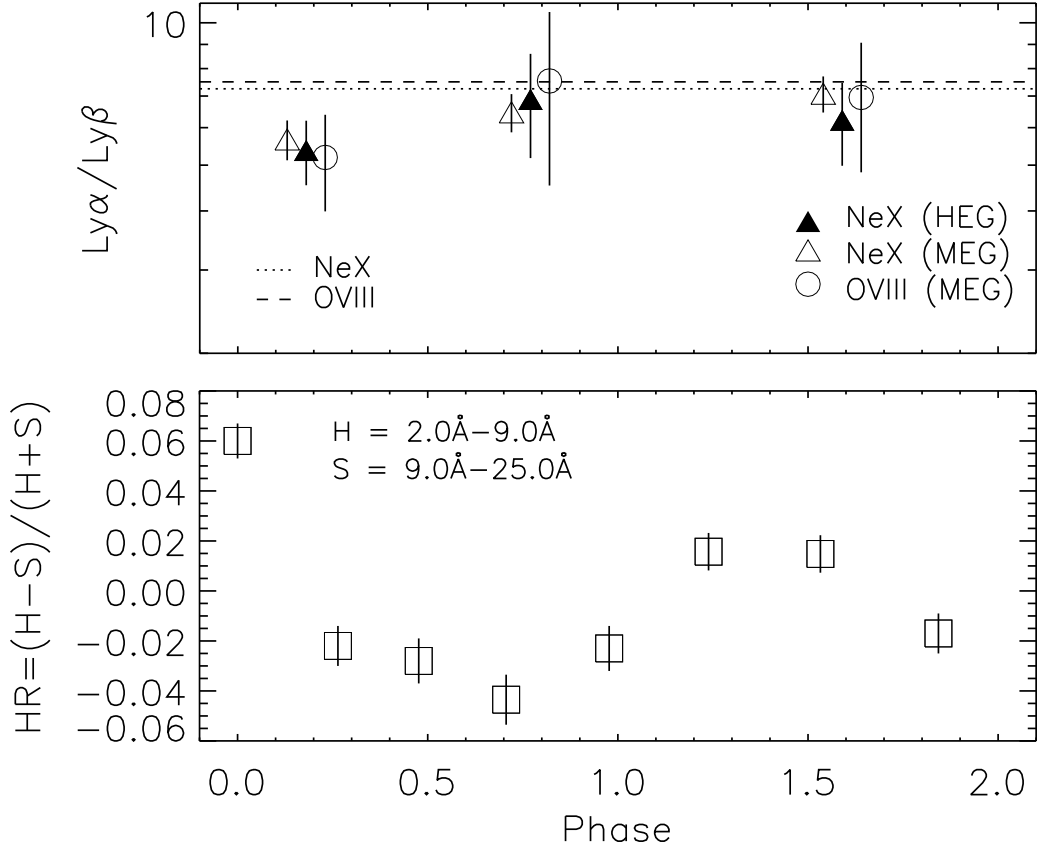}}\vspace{-0.8cm}
\caption{\lyab\ ratios measured from IM~Peg spectra for the three 
	time segments selected for the analysis (2528+2528, 
	2529+2530+2531, and 2532+2533+2534; {\em upper panel}), 
	together with the hardness ratio, HR ({\em lower panel}). 
	In the {\em upper panel} the horizontal lines represent the
	APED values expected for the \lyab\ ratio of O ({\em dashed 
	line}) for a temperature of 5~MK, and of Ne ({\em dotted 
	line}) for a temperature of 9~MK. For better readability 
	\lyab\ ratios of \nex\ from {\sc heg}, and of \oviii\, are 
	shifted with respect to {\sc meg} \nex\ \lyab\ ratios by 
	+0.05 and +0.1 respectively on the phase axis.
	\label{fig:Lyab_phLx}}
\end{figure}

The \lyab\ ratios, while statistically all consistent with one 
another, are suggestive of different conditions in the different
portions of the observation.  Both Ne and O \lyab\ ratios in 
{\sc heg} and {\sc meg} spectra are lower than the corresponding 
APED theoretical value in the first segment, as already discussed 
in Paper~I and noted earlier.
The rest of the observation is characterized by \lyab\ values
compatible with theory.  The measured ratios are illustrated as a
function of the hardness ratio ($HR=(H-S)/(H+S)$, where $H$ is the 
flux integrated in the 2-9\AA\ wavelength band, and $S$ is the flux
integrated in the 9-25\AA\ band) in the different segments of the 
observation in Figure~\ref{fig:Lyab_phLx}.  There is no obvious 
correlation of  \lyab\  with spectral hardness that might suggest, 
e.g., that the former changes as a result of plasma temperature 
changes. 
These results show that it might be worthwhile to explore,
when the data quality allows it, temporal variability of optical
depth properties in stellar coronae, expected to some extent on
the basis of their dependence on the line-of-sight and coronal
geometry other than on the physical conditions of the emitting plasma.

\clearpage
\clearpage

\subsection{Path length estimates}
\label{s:pathl}

The data presenting more reliable evidence for optical thickness are 
the spectra of IM~Peg from the first two observations, where \lya\ of 
both oxygen and neon appear to be depleted with respect to the 
corresponding \lyb\ transition (deviation of $\sim 1.8\sigma$ and 
$\sim 3 \sigma$ respectively), and the spectrum of II~Peg 
characterized by an  \oviii\ \lyab\ ratio $\sim 5 \sigma$ lower than 
the theoretical value.
As pointed out in \ta, the discrepancies between observed and 
theoretical \lyab\ line strength ratios in IM~Peg and II~Peg cannot be 
explained by photoelectric absorption along the line-of-sight on the 
basis of their measured H column densities \citep{Mewe97,Mitrou97}.  
It is worth discussing whether these transitions, and in 
particular their ratios, might be affected by 
non-equilibrium conditions that are possibly relevant for these 
active stars undergoing frequent flaring activity.
Non-equilibrium effects might be responsible for changes in the 
\ovii\ $G$~ratio ($(f+i)/r$) observed for 
EV~Lac between quiescent and flare phases 
\citep{Mitra-Kraev06}. It is well-known that 
the $G$~ratio is sensitive to deviations 
from coronal equilibrium, because recombination processes 
contribute significantly to $f$ and $i$ but are almost negligible for
$r$  (e.g.\ \citealt{Pradhan85}).
Even though recombination processes have non-negligible effects on
\lya\ and \lyb\ (10-15\%), they contribute to a very similar extent to 
the two transitions, and therefore their ratio is not expected to
change significantly under mildly non-equilibrium conditions.
Furthermore, we note that for the typical conditions of the coronal 
plasma in these active stars the expected timescales of these effects 
are only of the order of hundreds of seconds (e.g., 
\citealt{Golub89}), i.e.\ very short with respect to our integration 
times of several tens of kiloseconds.
We therefore interpreted the discrepant ratios in terms of a relative 
depletion of the \lya\ line flux due to resonance scattering processes.

Another source apparently showing significant effects of resonant
scattering is EV~Lac, whose \oviii\ \lyab\ ratio is more than
$3 \sigma$ lower than the corresponding theoretical value.
This result is supported to some extent by the findings of 
\citet{Ness03b}: in their survey of coronal optical depth 
properties based on the analysis of \fexvii\ transitions, 
EV~Lac is the only source deviating significantly from the
typical values found for all other coronae.  However, \citet{Ness03b}
did not consider this result robust due to discrepancies between 
{\sc heg} and {\sc meg} measurements.

For the stars with \lyab\ departing from the theoretical values we 
can derive an estimate of the photon path length using the 
\citet{Kaastra95} approximation to the escape probability formalism of
\citet{Kastner90}, as described in \S~3.1 of \ta.  The observed
\lyab\ can be expressed in terms of the line center optical depths
$\tau^k$, f-values $f^k$, and photon pathlength $\ell$, as 
\begin{equation} 
\frac{(I_{Ly\alpha}/I_{Ly\beta})_{\mathrm{obs}}}
{(I_{Ly\alpha}/I_{Ly\beta})_{\mathrm{th}}} =
   \frac{1}{3} \left[\frac{2}{1+0.43 C(\ell) f^{\alpha}_1}
    + \frac{1}{1+0.43 C(\ell) f^{\alpha}_1/2}\right]
    \cdot (1+0.43 C(\ell) f^{\beta})
\end{equation}
where $C(\ell) \cdot f^k =\tau^k $ and the line center optical depth
is given by 
\begin{equation}
 \tau = 1.16 \cdot 10^{-14} \cdot
     \frac{n_{\mathrm{i}}}{n_{\mathrm{el}}} A_{\mathrm{Z}}
    \frac{n_{\mathrm{H}}}{n_{\mathrm{e}}}
    \lambda f \sqrt{\frac{M}{T}} n_{\mathrm{e}} \ell 
    \label{eq:tau}
\end{equation}
where $n_{\mathrm{i}}/n_{\mathrm{el}}$ is the ion fraction 
\citep[from][]{Mazzotta98}, $A_{\mathrm{Z}}$ is the element abundance,
$n_{\mathrm{H}}/n_{\mathrm{e}} \sim 0.85$, $f$ is the oscillator 
strength, $M$ the atomic weight, $T$ the temperature, and 
$n_{\mathrm{e}}$ the electron density. The \lya\ and \lyb\ f-values 
are $f^{\alpha}_1 = 0.2776$, and $f^{\beta}_1 = 0.05274$ (e.g.\ 
\citealt{Morton03}).

We use the electron densities, $n_{\mathrm{e}}$, derived from the 
diagnostics of the He-like triplets \tb.  
The coronal abundances assumed for II~Peg and IM~Peg are discussed in
\ta; for EV~Lac we assume O/H=8.40, as derived by \citet{Favata00c}, 
expressed on the usual spectroscopic logarithmic scale in which 
X/H$= \log [n({\rm X})/n({\rm H})]+12$, where $n({\rm X})$
is the number density of element X.

In Table~\ref{tab:pathl} we list the values obtained for the path 
length estimates, $\ell_{\tau}$, and for comparison, we list the 
stellar radii and loop lengths expected for a standard hydrostatic 
loop model (e.g.\ \citealt{RTV}, RTV hereafter) corresponding to the 
observed temperatures and densities.

\clearpage
\begin{deluxetable}{lcccclll}
\tabletypesize{\footnotesize}
\tablecaption{Path length derived from measured \lyab\ ratios.
	\label{tab:pathl}}
\tablewidth{0pt}
\tablehead{
 \colhead{Source} & \colhead{} & \colhead{} & \colhead{$\ell_{\tau}$} 
 & \colhead{$L_{\mathrm{RTV}}$ \tablenotemark{a}}
 & \colhead{$\ell_{\tau}$/R$_{\star}$ \tablenotemark{b}}
 & \colhead{$f_s$ \tablenotemark{c}} & \colhead{$E_S$ \tablenotemark{d}}  \\
 \colhead{} & \colhead{} & \colhead{} & \colhead{[cm]} 
 &  \colhead{[cm]} & \colhead{} & \colhead{} 
 & \colhead{[erg~cm$^{-2}$~s$^{-1}$]}
}
\startdata
IM~Peg & \oviii\ &       & $1.5 \cdot 10^{10}$ & $2.2 \cdot 10^{9}$ & 0.017   & $\sim 0.006$  & $\sim 7.2 \cdot 10^7$ \\  
       & \nex\   & [HEG] &  $2.1 \cdot 10^{8}$ & $2.8 \cdot 10^{7}$ & 0.00024 & $\sim 0.0003$ & $\sim 5.2 \cdot 10^{10}$\\  
       & \nex\   & [MEG] &  $1.7 \cdot 10^{8}$ & $2.8 \cdot 10^{7}$ & 0.00019 & $\sim 0.0004$ & $\sim 5.3 \cdot 10^{10}$\\[0.15cm] 
II~Peg & \oviii\ &       &  $9.5 \cdot 10^{9}$ &   $1 \cdot 10^{9}$ & 0.04    & $\sim 0.021$  & $\sim 1.2 \cdot 10^8$\\[0.15cm]
EV~Lac & \oviii\ &       &  $1.6 \cdot 10^{9}$ & $2.6 \cdot 10^{8}$ & 0.058   & $\sim 0.016$  & $\sim 1.9 \cdot 10^8$\\[0.15cm]
 \enddata 
\tablenotetext{a}{Loop length from RTV scaling laws:
   	$L_{\mathrm{RTV}} \sim T^3/[(1.4 \times 10^3)^3 \cdot p]$.}
\tablenotetext{b}{Path length as fraction of the 
   	stellar radius.}
\tablenotetext{c}{Coronal surface filling factor, defined 
	as $f_s=A/A_{\star}=(V/\ell_{\tau})/A_{\star}
	=[EM/(n_{\mathrm{e}}^2 \ell_{\tau})]/A_{\star}$.}
\tablenotetext{d}{Estimate of surface heating flux 
	(see text for details).}
\end{deluxetable}

\clearpage
\section{Discussion}
\label{s:discuss}

This work present a detailed and extensive study of the \oviii\ and 
\nex\ \lyab\ ratios in a sample of active stars, paying careful 
attention to the presence of blends and to the validity of theoretical 
line ratio predictions. Our analysis shows that optical depth effects 
are generally negligible in the disk-integrated X-ray spectra emitted 
by stellar coronae over a wide range of activity, in line with previous 
studies based on \fexvii\ lines \citep[e.g.,][]{Ness03b,Audard04}.  
We argued in \S\ref{s:theor} that the coronal abundance patterns 
exhibited by most active stars renders the Ne and O Lyman lines more
sensitive diagnostics of optical depth than the Fe lines, and in this
regard our study extends the results of previous surveys.

However, our study has also yielded convincing indication of 
line-of-sight photon loss through resonant scattering in the spectra 
of three stars, IM~Peg, II~Peg and EV~Lac.  The detection of 
significant X-ray optical depth is particularly interesting, because 
it provides us with insights into the characteristic dimensions of 
the emitting coronal structures.  

Before proceeding, we note here that a simple escape probability
analysis does not include the scattering source term itself, and so
photons scattered {\em into} the line of sight that could enhance the
observed line strength are not included.  For instance, in the case of
a spherically symmetric corona in which the line optical depth were
significant, scattering into and out of the line of sight would be
balanced and line strengths not affected (see e.g., \citealt{Wood00}).  
Strictly, then, the photon path, $\ell$, entering into the equations 
in \S\ref{s:pathl} should be interpreted as a {\em lower limit} to 
the true photon path length.   This is potentially important.  
\citet{Phillips01}, for example, used the lack of appreciable 
depletion of the \fexvii\ resonant line at $\sim 15.01$\AA\ to deduce
an upper limit of 3000~km for size of emitting regions in the corona 
of Capella; however, such a measurement in reality only provides an 
{\em upper limit to the lower limit}, therefore not providing a 
constraint.  
Indeed, the lack of evidence for line quenching through scattering in 
most of our sample stars does not imply that scattering is not a 
significant source term, but merely that the scattering geometry is such
that there is no net loss or gain of photons in the line-of-sight. 

Instead, a positive detection of resonance scattering has interesting
implications and points to a non-uniform ``aspect ratio'' of the
dominant coronal emitting regions: for any net line depletion to
occur, an emitting structure must generally be more elongated along
the line of sight than in the perpendicular direction.  In the context
of a corona comprising plasma contained by magnetic loops, there are
different ways in which this can be interpreted: (1) scattering
loss results from the structure and viewing angle of the loops
themselves; (2) scattering loss arises because of a particular
conglomeration of loops viewed at a particular angle.  Each case leads
to fundamentally different interpretations of observed scattering.  In
the former case, a loop with random orientation must generally be
viewed from the top, in which direction the photon path length through
the plasma is largest.  Photon loss through resonance scattering then
implies that loops are preferentially placed on the stellar surface
facing the observer (a perfect alignment of loops seen edge-on on the
stellar limb could produce a similar effect, though such a chance
alignment is unlikely).  In the latter case, the coronal structures or
active regions must be placed preferentially on the stellar limb.

\subsection{Scattering within individual loops}

With the proviso that a scattering-derived path length is a lower limit
to the true emitting region size, we note that the photon path lengths 
implied by the analysis of line ratios (\S\ref{s:pathl}) observed in 
II~Peg, IM~Peg and EV~Lac and listed in Table~\ref{tab:pathl} are all
much smaller than the corresponding stellar radii, but are also about 
an order of magnitude {\em larger} than expected from RTV loops.

Under the assumption that scattering within individual loops arises
when they are viewed from the top, we can interpret the path length
estimates, $\ell_{\tau}$, in terms of the coronal scale height.  With
knowledge of the total emitting volume we can also derive an estimate
of the surface filling factors.  Emitting volumes can be estimated
using the emission measures implied by the different lines,
$V=EM/n_{\mathrm{e}}^2$, combined with electron densities,
$n_{\mathrm{e}}$, derived in \tb.  The surface filling factors, $f_s$,
are then given by $f_s=A/A_{\star}=(V/\ell_{\tau})/A_{\star}$.  The
derived filling factors, listed in Table~\ref{tab:pathl}, are very
small, and especially so for the hotter plasma.  They are also an
order of magnitude smaller than those previously derived in \tb\ based
on RTV loops---a direct consequence of the optical depth scale height
being commensurately larger than those suggested by simple
quasi-static loop models.

The small filling factors we find here have interesting implications
for the surface heating flux requirements.  A very rough estimate of
the heating required to sustain the observed physical conditions of
the plasma can be obtained from the RTV scaling laws.  While our
estimated loop lengths are significantly longer than suggested by RTV
models, these relations should still suffice for the purposes of
estimation.  The volumetric heating per unit time, $E$, is given by $E
\sim 10^5 \cdot p^{7/6} L^{-5/6}$, where $p$ is the plasma pressure
and $L$ the loop length.  By assuming $L = \ell_{\tau}$ we find for
the the surface flux values of order of $10^8$~erg~cm$^{-2}$~s$^{-1}$
for \oviii\ lines, and several $10^{10}$~erg~cm$^{-2}$~s$^{-1}$
implied by the \nex\ lines in IM~Peg (Table~\ref{tab:pathl}).  These
compare with typical surface heating rates for the cores of solar
active regions of a few $10^7$~erg~cm$^{-2}$~s$^{-1}$
\citep[e.g.,][]{Withbroe77}.

In the comparison of $\ell_{\tau}$ with $L_{\rm RTV}$ we assumed to
some extent that $\ell_{\tau}$ is a reasonable estimate for the
coronal scale height (i.e.\ for the loop length).  However, an 
alternative interpretation is possible in which the actual coronal 
loops are larger and $\ell_{\tau}$ represents only an estimate of the 
length of the part of the loop containing plasma at the characteristic
temperature of the quenched lines.  In such a scenario, the values
found for $\ell_{\tau}$ can suggest the presence of cooler loops with
maximum temperature around 3~MK, similar to solar loops, coexistent
with loops with much higher maximum temperature ($\gtrsim$10~MK).  In
these hotter loops, the plasma emitting the \nex\ lines ($T \gtrsim$
6-7~MK) would be confined in the lower portion of the loop, which is
characterized by higher density.  In order to estimate the region of
the loop occupied by the plasma emitting the observed line,
$\Delta l$, we can use the equations for the conductive flux:
\begin{equation}
 	F_c = k_c T^{5/2} \cdot dT/dl \sim 
	k_c T^{5/2} \cdot \Delta T/\Delta l ;
\end{equation}
and from RTV relations for a radiative power loss function $P(T)$:
\begin{equation}
	F_c / \Delta l \sim n_{\mathrm{e}}^2 P(T).
\end{equation}
Assuming T is the temperature of maximum formation of the line $k$, 
$T^{k}_{\mathrm{max}}$, and $\Delta T$ is the width in temperature 
of the line emissivity curve ($\sim 0.3$~dex), we can solve the 
equations for $\Delta l$:
\[
\Delta l \sim 5 \cdot 10^8 \mbox{cm   for \oviii}
\]
\[
\Delta l \sim 2 \cdot 10^7 \mbox{cm   for \nex}
\]
The resulting values of $\Delta l$ are rather close to $L_{RTV}$ 
(see Table~\ref{tab:pathl}), obtained assuming $T^{k}_{\mathrm{max}}$ 
as maximum temperature of the loop, and are still much smaller than
$\ell_{\tau}$.  We can conclude then, that the values derived for the 
path length do not seem to agree with the hypothesis of standard
uniformly heated quasi-static loop models.

\subsection{Scattering within active regions}

The maximum path length, $l_{\rm max}$, through a spherically-symmetric 
corona of height $h$ on a star with radius $R_\star$ and surface 
filling factor $f_s$ (assuming this filling factor does not vary 
significantly with height) is
\begin{equation}
l_{\rm max}=2f_s\sqrt{h^2+2hR_\star}.
\end{equation}

Considering typical parameters found for stellar coronae, we can 
estimate from the above equation whether or not we expect significant 
photon scattering in stellar coronae, regardless of whether we see a 
net photon loss.  From Eq.~\ref{eq:tau} we can estimate the optical 
depth at the limb for Ne and O \lya\ lines.
Assuming the temperature of maximum formation of the line, i.e.\ about
6~MK and 3~MK for Ne and O respectively, and typical density of about
$10^{10}$~cm$^{-3}$ and $10^{11}$~cm$^{-3}$ respectively (see e.g., 
\tb, \citealt{Ness04}), we derive
$\tau({\rm Ne}) \sim 5.2\times 10^{-10} \cdot l_{\rm max}$ and 
$\tau({\rm O}) \sim 3.6\times 10^{-10} \cdot l_{\rm max}$. 

The height derived for coronal structures with different techniques
(see \S\ref{s:intro} for references) is $\lesssim R_\star/2$;
on the Sun a typical coronal height is closer to $R_\star/10$.
Assuming $h = R_\star/10$, and $R_\star =R_\odot$ we obtain 
$l_{\rm max} \sim f_s \cdot 0.9 R_\star \sim f_s \cdot 6 \times 10^{10}$~cm.

Under these assumptions, the values of the optical depth at the limb, 
for a spherically symmetric corona, are therefore 
$\tau({\rm Ne}) \sim 30f_s$ and $\tau({\rm O}) \sim 20f_s$ (or 
$\tau({\rm Ne}) \sim 80f_s$ and $\tau({\rm O}) \sim 55f_s$ if we 
assume $h = R_\star/2$, and $R_\star =R_\odot$). 
For the typical filling factors lower than a few percent (e.g., \tb) 
we obtain optical depth $\tau \lesssim 1$ both for Ne and for O.

\subsection{Correlation of scattering with stellar and coronal 
parameters} 

While only three of our stellar sample present a significant case for
resonance scattering, it is possible that trends of the departure from
the theoretical \lyab\ ratio with fundamental stellar parameters can
be found.  We have examined the departures of observed from
theoretical \lyab\ ratios as a function of X-ray surface flux, surface
flux in the \lya\ line, \lx/\lbol, filling factors, and plasma
density.  These comparisons are suggestive of correlations
between quenching of \lya\ photons and both \lx/\lbol\ and the
density-sensitive ratio of strengths of the forbidden and
intercombination lines, $f/i$, of \mgxi.

\begin{figure}[!h]
\centerline{\includegraphics[width=8cm]{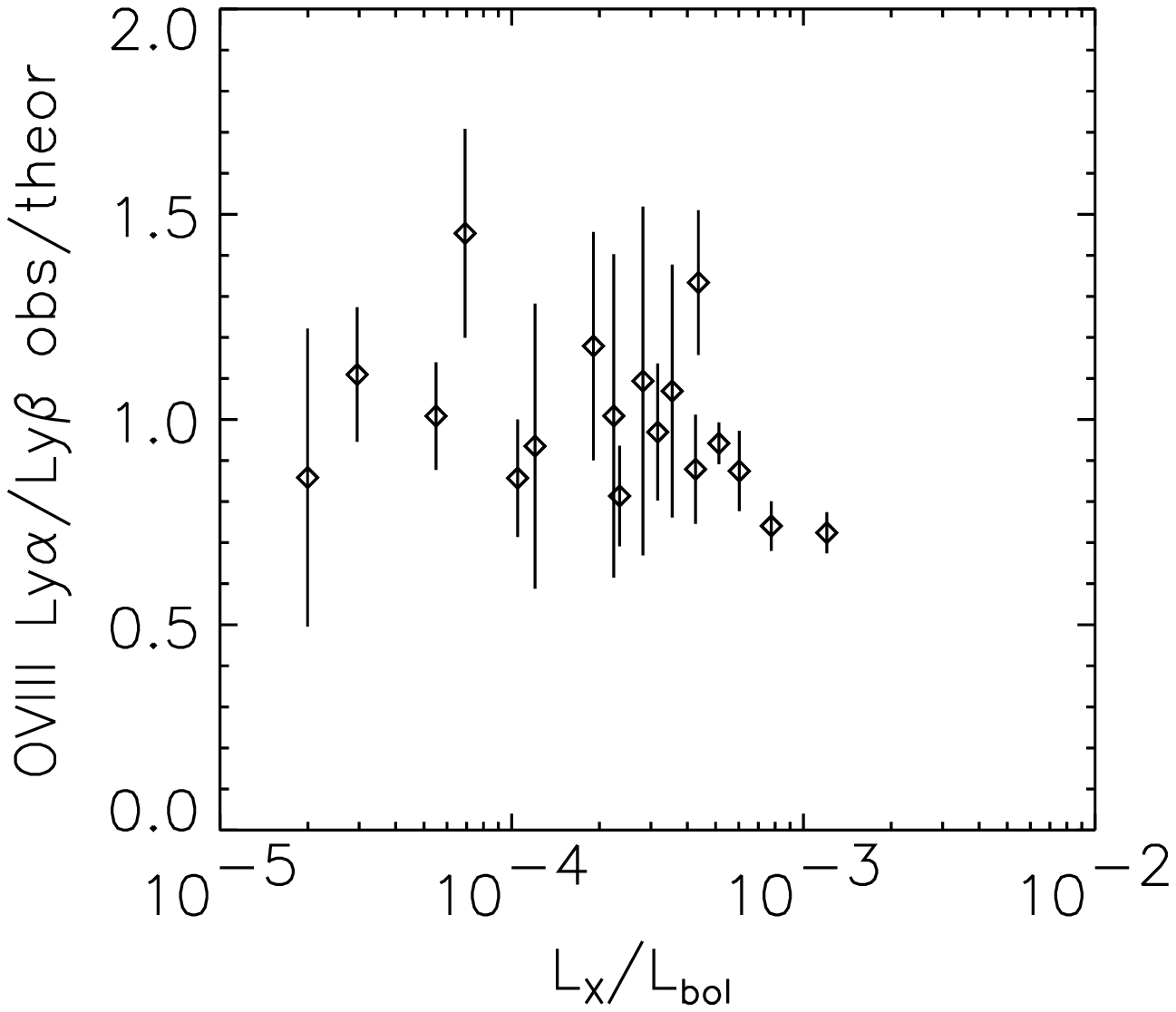}}\vspace{-0.5cm}
\centerline{\includegraphics[width=8cm]{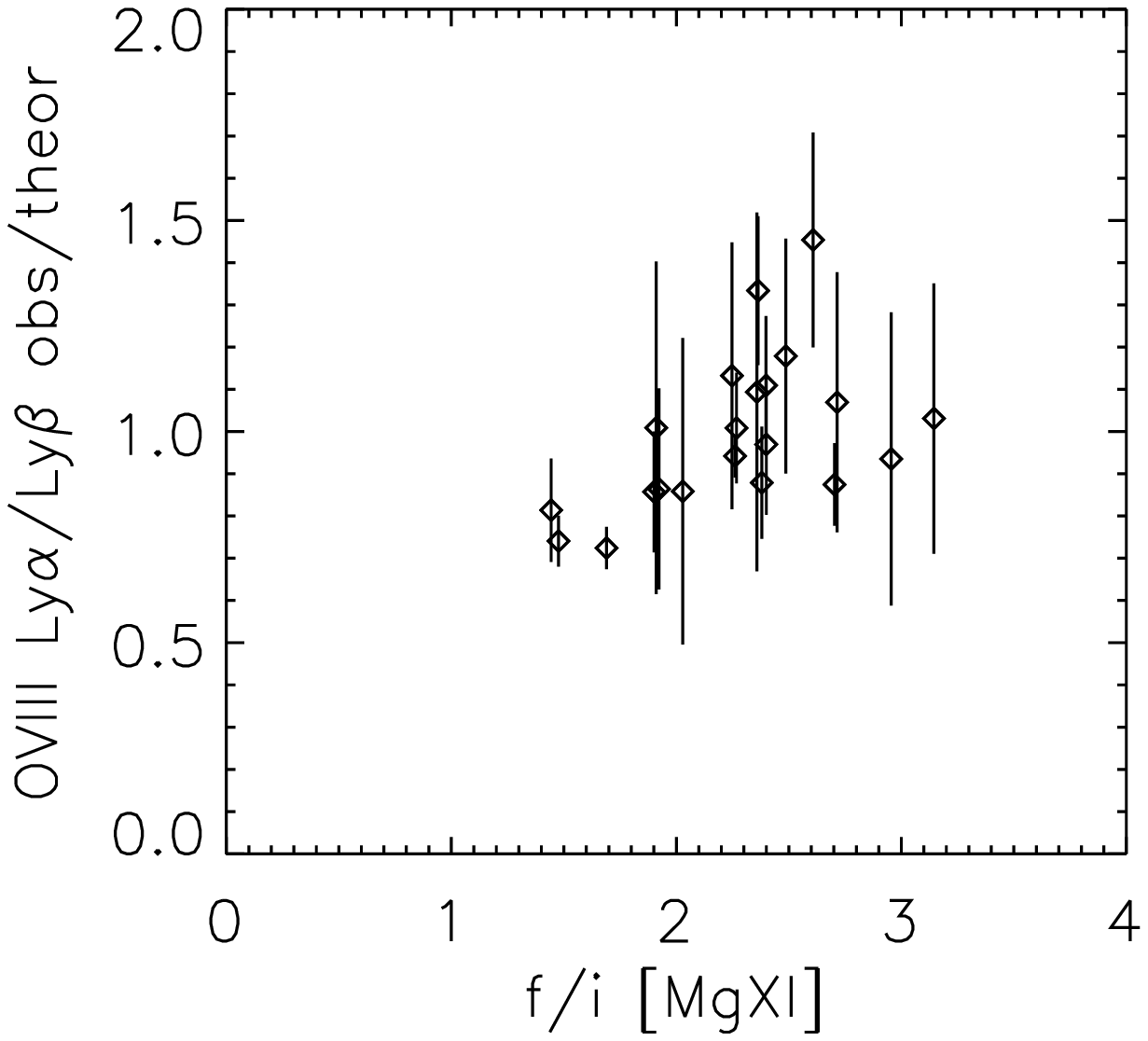}}
\caption{Ratios of measured to theoretical \oviii\ \lyab\ ratios
	plotted vs.\ \lx/\lbol\ ({\em top}) and vs.\ the \mgxi\ 
	forbidden to intercombination line ratio (from \tb; 
	{\em bottom}), which is a diagnostic of plasma density 
	(low f/i ratios correspond to high density).
	  \label{fig:trends}}
\end{figure}

These correlations are illustrated in Figure~\ref{fig:trends} where we
show the measured to theoretical ratios of \oviii\ \lyab\ of our
sample as a function of \lx/\lbol\ ({\em top}) and \mgxi\ $f/i$ ({\em
bottom}).  Error-weighted linear fits to the data in these figures
yield slopes of $-0.11\pm 0.04$ and $0.25\pm 0.06$, respectively.
The sources presenting evidence of optical depth effects are the stars
characterised by the highest activity level (\lx/\lbol) and the
highest plasma densities (i.e.\ lowest f/i) in our sample.  Such
correlations are what are expected based on Eqn.~\ref{eq:tau}.
Optical depth is proportional to the product of electron density and
typical path length within an emitting region, $n_e \ell$.  For a
given fixed volume emission measure, $n_e^2 V \sim n_e^2 l^3$, the
optical depth varies as $n_e^{1/3}$ or $\ell^{-1/2}$, and so increases
with increasing plasma density.  In the case of \lx/\lbol, an increase
in \lx\ can arise through either $\ell$ or $n_e$, such
that any increase in \lx\ might typically be expected to lead to
greater scattering optical depth.  While only a small handful of our
measurements present truly significant detections of resonance
scattering, the evidence for trends of increasing $\tau$ with
\lx/\lbol\ and \mgxi\ $f/i$ as is expected adds confidence to the
interpretation of the \lyab\ ratios in these terms.

\clearpage
\clearpage

\section{Conclusions}
\label{s:conclude}

We have investigated the optical thickness of stellar coronae through
the analysis of \lya\ and \lyb\ lines of hydrogen-like oxygen and neon
ions, in \cha-\hetg\ spectra of a large sample of active stars.  Our
study indicates that most stellar coronae are characterised by
negligible visible signs of optical depth, in agreement with the
results of previous studies based on \fexvii\ lines.  This indicates
that coronae are either in general optically-thin, or that for cases
in which optical depths reach of order unity or higher the 
geometry does not strongly favour lines-of-sight showing net \lya\
photon loss.

We do find evidence of significant optical depth in the \oviii\ Lyman
lines of the RS CVn binary II~Peg, and of the single M dwarf EV~Lac;
the RS CVn binary IM~Peg also shows depletion of both Ne and O \lyab\
ratios as compared with theoretical predictions, and our analysis
indicates that it is a transient effect present only in part of the
observations.  The detection of significant optical depth allows to
derive an estimate for the photon path length and therefore for the
typical height of the corona.  The size of coronal structures derived
for all three sources is of the order of a few percent of the stellar
radius at most, implying very small coronal filling factors and high
surface heating fluxes.  We searched for correlation with basic
stellar parameters and coronal properties and we find that the sources
presenting evidence of significant optical depth are at the high end
of activity level, with \lx/\lbol\ at the saturation limit, and high
densities in their hot plasma, as revealed by the \mgxi\ He-like
triplet lines.

For the stellar sample as a whole, we also find evidence of increasing
quenching of \lya\ relative to \lyb\ as function of both \lx/\lbol\ and
the density-sensitive \mgxi\ forbidden to intercombination line ratio.
Such a trend is expected in the scenario in which optical depths are
significant but generally small: viewing geometry rarely favours large
net photon enhacements, but for favourable lines-of-sight photon
depletion is expected to increase with both increasing \lx\ and
increasing plasma density.

\acknowledgements
PT and DH were supported by SAO contract SV3-73016 to MIT for support 
of the {\em Chandra X-ray Center}, which is operated by SAO for and 
on behalf of NASA under contract NAS8-03060.  JJD was supported by 
NASA contract NAS8-39073 to the {\em Chandra X-ray Center} during 
the course of this research.  GP acknowledges support from Agenzia 
Spaziale Italiana and italian Ministero dell'Universit\`a e della 
Ricerca.

\end{document}